\documentclass[superscriptaddress,preprint,amsmath,amssymb,aip,longbibliography,floatfix]{revtex4-2}

\usepackage{xcolor}
\usepackage{graphicx}
\usepackage{dcolumn}
\usepackage{bm}
\usepackage[normalem]{ulem} 
\usepackage[english]{babel}
\usepackage{hyperref}
\hypersetup{
	colorlinks,
	citecolor=blue,
	filecolor=black,
	linkcolor=black,
	urlcolor=black
}

\begin{document}
\selectlanguage{english}

\title{Structural, magnetic, and nanoacoustic characterization of Co/Pt superlattices}%

\author{E. R. Cardozo de Oliveira}
\affiliation{
Université Paris-Saclay, C.N.R.S., Centre de Nanosciences et de Nanotechnologies (C2N),  10 Boulevard Thomas Gobert, 91120 Palaiseau, France
}

\author{C. Xiang}
\affiliation{
Université Paris-Saclay, C.N.R.S., Centre de Nanosciences et de Nanotechnologies (C2N),  10 Boulevard Thomas Gobert, 91120 Palaiseau, France
}

\author{C. Borrazás}
\affiliation{
Universidad de Buenos Aires, Facultad de Ciencias Exactas y Naturales, Departamento de Física, 1428 Buenos Aires, Argentina
}

\author{S. Sandeep}
\affiliation{
Université Paris-Saclay, C.N.R.S., Centre de Nanosciences et de Nanotechnologies (C2N),  10 Boulevard Thomas Gobert, 91120 Palaiseau, France
}

\author{J. E. Gómez}
\affiliation{
Instituto de Nanociencia y Nanotecnología, CNEA-CONICET, Centro Atómico Bariloche, Av. E. Bustillo 9500, R8402AGP San Carlos de Bariloche, Río Negro, Argentina
}

\author{M. Vásquez Mansilla}
\affiliation{
Instituto de Nanociencia y Nanotecnología, CNEA-CONICET, Centro Atómico Bariloche, Av. E. Bustillo 9500, R8402AGP San Carlos de Bariloche, Río Negro, Argentina
}

\author{N. Findling}
\affiliation{
Université Paris-Saclay, C.N.R.S., Centre de Nanosciences et de Nanotechnologies (C2N),  10 Boulevard Thomas Gobert, 91120 Palaiseau, France
}

\author{L. Largeau}
\affiliation{
Université Paris-Saclay, C.N.R.S., Centre de Nanosciences et de Nanotechnologies (C2N),  10 Boulevard Thomas Gobert, 91120 Palaiseau, France
}

\author{N. D. Lanzillotti-Kimura}
\email{daniel.kimura@c2n.upsaclay.fr}
\affiliation{
Université Paris-Saclay, C.N.R.S., Centre de Nanosciences et de Nanotechnologies (C2N),  10 Boulevard Thomas Gobert, 91120 Palaiseau, France
}

\author{M. Granada}
\email{maragranada@cnea.gob.ar}
\affiliation{
Instituto de Nanociencia y Nanotecnología, CNEA-CONICET, Centro Atómico Bariloche, Av. E. Bustillo 9500, R8402AGP San Carlos de Bariloche, Río Negro, Argentina
}

\affiliation{
Instituto Balseiro, Universidad Nacional de Cuyo - CNEA, Av. E. Bustillo 9500, R8402AGP San Carlos de Bariloche, Río Negro, Argentina
}
\date{\today}

\begin{abstract}
Superlattices presenting a spatial modulation of the elastic properties appear as a main tool to reach the THz regime in nanoacoustic devices. The exploration of alternative materials with multifunctional properties remains a fertile domain of research.  In this work, we study the structural, magnetic, and
acoustic characteristics of nanometric superlattices made of Pt/Co. 
The samples present a well defined periodicity, as determined by X-ray reflectometry, whereas scanning transmission electron microscopy with local compositional analysis reveals that the superlattices present a modulation in composition instead of sharp interfaces. The policrystalline nature of the superlattices is evidenced both by X ray diffraction and transmission electron microscopy.
Magnetization measurements show a perpendicular magnetic anisotropy for the higher Co concentrations. 
Picosecond acoustic experiments evidence that the studied samples support short-lived acoustic modes up to 900 GHz, and up to 7 acoustic echoes at lower frequencies.
These are promising results for the development of magnetoacoustic devices working at ultrahigh frequencies.

\end{abstract}

\maketitle

\section{\label{introduction}Introduction}

Ferromagnetic materials, such as Fe, Ni, Co and their alloys, have formed the basis for data storage and processing devices.\cite{comstockReviewModernMagnetic2002} Magnetism in these materials is usually controlled through an external magnetic field, voltage, electric field or light.\cite{beaurepaireUltrafastSpinDynamics1996,bigotFemtosecondSpectrotemporalMagnetooptics2004,weisheitElectricFieldInducedModification2007,chuElectricfieldControlLocal2008,ranaMagnonicDevicesBased2019} Recently, acoustic waves have emerged as a novel candidate to manipulate magnetism and spin dynamics via phonon-spin coupling and phonon-magnon coupling, which holds great potential for applications in more compact and faster devices.\cite{kittelInteractionSpinWaves1958,liAdvancesCoherentCoupling2021,yangAcousticControlMagnetism2021,hiokiCoherentOscillationPhonons2022,weberEmergingSpinPhonon2022,kalashnikovaImpulsiveGenerationCoherent2007,scherbakovCoherentMagnetizationPrecession2010,thevenardEffectPicosecondStrain2010,kuszewskiOpticalProbingRayleigh2018,thevenardPrecessionalMagnetizationSwitching2016} In this context, characterizing magnetic and acoustic properties is fundamental to understanding and exploring these mechanisms. 

Metallic ferromagnetic Co/Pt multilayers present perpendicular magnetic anisotropy which can be tailored by adjusting the total structure thickness.\cite{hashimotoCoPtCo1990,hashimotoFilmThicknessDependence1990} Previous studies have reported that the acoustic phonon modes in such structures can also be engineered by leveraging electron-lattice interactions that are dependent on the multilayer repeat number.\cite{shimUltrafastGiantMagnetic2017,kimCoherentPhononControl2016} While most works focus on the gigahertz (GHz) acoustic phonons confined within the entire structure, few works have investigated the ultrahigh-frequency acoustic phonons in magnetically ordered superlattices (SLs), close to terahertz (THz) frequency range.\cite{guSuperordersTerahertzAcoustic2024} 

Nanometer-scale superlattices enable the engineering of the acoustic phonon dispersion relation and dynamics. The additional periodicity introduced by the SLs induces the reduction of the Brillouin zone, and thus the folding of the acoustic branches, and the opening of acoustic minigaps at the Brillouin zone edge and center.\cite{huynhSemiconductorSuperlatticesTool2015,lanzillotti-kimuraNanowaveDevicesTerahertz2006,lanzillotti-kimuraAcousticPhononNanowave2007,narayanamurtiSelectiveTransmissionHighFrequency1979} The thickness of the superlattice period determines the frequencies at which the minigaps appear, while the composition and relative thickness of the materials composing the unit cell set the gap bandwidth.\cite{ortizTopologicalOpticalPhononic2021,ortizPhononEngineeringSuperlattices2019} For instance, acoustic SLs can be used as phonon filters, or as phonon mirrors in a phononic Fabry-Perot resonator. They can constitute the building blocks of complex heterostructures able to control both hypersound and light dynamics, and can be used to mimic solid state physics phenomena, e.g. topological states, Bloch oscillations and Anderson localization, to name a few.\cite{arreguiCoherentGenerationDetection2019,esmannTopologicalNanophononicStates2018,ortizTopologicalOpticalPhononic2021,rodriguezTopologicalNanophononicInterface2023,lanzillotti-kimuraBlochOscillationsTHz2010,arreguiAndersonPhotonPhononColocalization2019}

While most of the reported works are based on III-V semiconductors, phononic SLs have been also reported in epitaxially grown oxides,\cite{lanzillotti-kimuraEnhancementInhibitionCoherent2010} polymers,\cite{gomopoulosOneDimensionalHypersonicPhononic2010a,schneiderDefectControlledHypersoundPropagation2013} and group IV semiconductors,\cite{ezzahriCoherentPhononsSi2007,wilsonEvidenceTerahertzAcoustic2018,parsonsBrillouinScatteringPorous2012} among others.\cite{perrinPicosecondUltrasonicsStudy1996,cardozodeoliveiraDesignCosteffectiveEnvironmentresponsive2023} The seek for new materials to engineer superlattices has become an asset for advancing the field of nanophononics, enabling innovative functionalities. For instance, responsive nanophononic devices with dynamical tuning of the acoustic response under external control are emerging as promising area of research.~\cite{priyaPerspectivesHighfrequencyNanomechanics2023} In this work, we study the structural, magnetic, and acoustic characteristics of ferromagnetic superlattices based on platinum (Pt) and cobalt (Co). We demonstrate that these SLs can support ultrahigh-frequency acoustic modes, close to 900 GHz. These results constitute a step forward in achieving ultrafast dynamics of phonon-magnon interactions.

The article is organized as follows: Sect. II presents a description of the fabricated samples as well as their structural and chemical characterization through X-ray reflectometry and scanning transmission electron microscopy. The magnetic response of the superlattices is discussed in Sect. III. Sect. IV is devoted to the coherent acoustic phonon dynamics. Finally, in Sect. V we discuss the results and propose some perspectives. 

\section{Structural characterization} \label{structural}

Co/Pt superlattices (SLs) were deposited by magnetron sputtering on (001) Si substrates with a native oxide layer, under argon pressure $P_\text{Ar} = 2.8$ mTorr as described elsewhere.\cite{QuinterosAPL}
The [Pt($t_{\text{Pt}}$)/Co($t_{\text{Co}}$)] period was repeated 10 times and a Pt($t_{\text{Pt}}$) capping layer was deposited on top. The deposition time for Co was varied throughout the series of samples, while the deposition time for Pt was kept constant to get $t_{\text{Pt}} = 4.2$ nm.

X-ray reflectometry (XRR) measurements were performed using a XPert PANalytical system equipped with a Cu anode in a sealed tube, with parallel plate collimators on both the incident and receiving paths. 
\begin{figure}[h]
    \includegraphics[width=0.7\columnwidth]{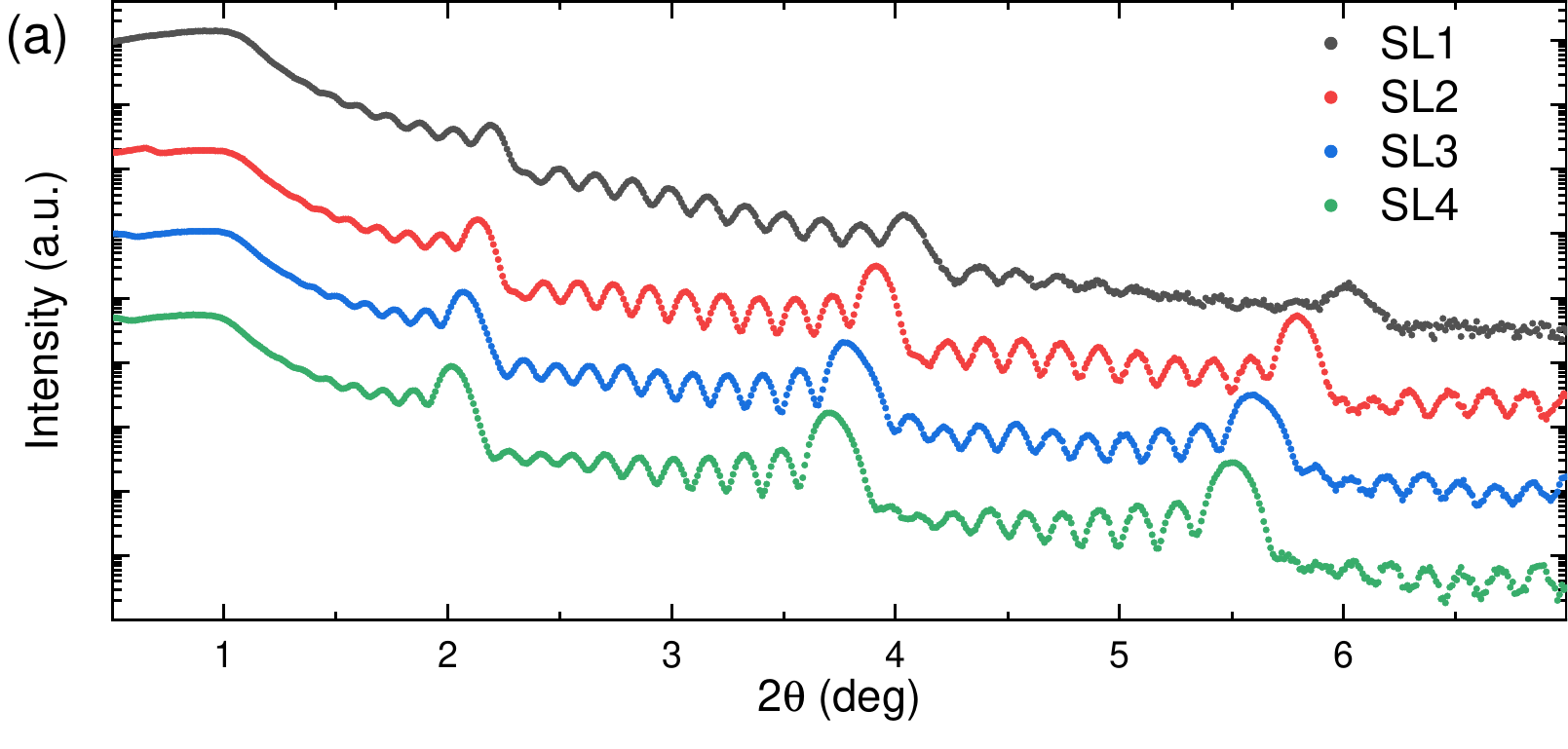}
    \includegraphics[width=0.7\columnwidth]{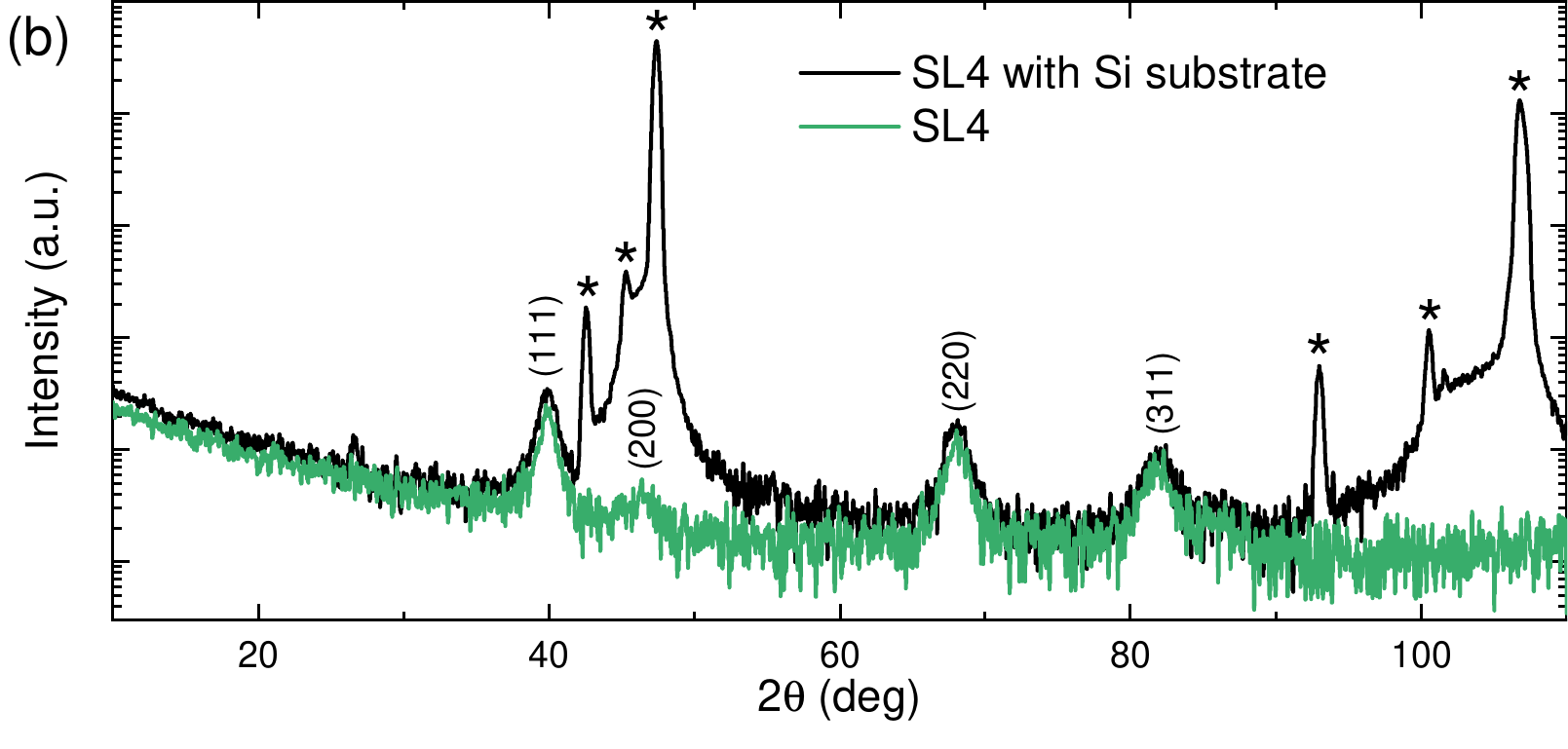}
    \caption{\footnotesize (a) X-ray reflectometry of the superlattices [Pt($t_{\text{Pt}}$)/Co($t_{\text{Co}}$)]$_{10}$/Pt($t_{\text{Pt}}$) studied in this work, measured with Cu-K$\alpha$ radiation. (b) Non coplanar grazing incidence X ray diffraction patterns. The black curve was obtained with the substrate in the Bragg condition, the Si diffraction peaks are indicated (*). The green curve was obtained with the substrate out of the Bragg condition, which confirms that the remaining peaks are associated to the SL. Those peaks are labeled with the Miller indices of the conventional cubic unit cell for Pt.}
    \label{fig.Xrays}
\end{figure}

They show clear superlattice peaks along with the total thickness fringes (see Fig. \ref{fig.Xrays}(a)). It was not possible to fit exactly the experimental curves. 
From the analysis of the positions of both sets of peaks, and assuming pure Pt and Co layers, we obtained the average values $t_{\text{Pt}} = (4.2 \pm 0.1)$ nm, and a deposition rate for Co of ($1.8 \pm 0.6$) nm/min, yielding estimated Co thicknesses $t_{\text{Co}}$ between 0.18 and 0.72 nm along the series of SLs. The samples studied in this work are: SL1 ($t_{\text{Co}}= 0.18$ nm), SL2 ($t_{\text{Co}}= 0.45$ nm), SL3 ($t_{\text{Co}}= 0.54$ nm) and SL4 ($t_{\text{Co}}= 0.72$ nm). Complementary studies by X-ray diffraction (XRD) and transmission electron microscopy (TEM) with energy dispersive spectroscopy (EDS) were performed on the sample with the largest total thickness, SL4.

The XRD measurements were performed using a Rigaku Smartlab diffractometer equipped with a rotating Cu anode and a five-circle goniometer.
XRD patterns in the conventional Bragg-Brentano configuration did not show any peaks corresponding to the SL, only intense peaks associated to the Si substrate were observed. 
Thanks to the 5-circle goniometer, we performed non coplanar grazing incidence X-ray diffraction experiments, which allow the crystallographic planes perpendicular to the surface to be put into diffraction conditions. This grazing incidence configuration makes it possible to exacerbate the diffraction of materials on the surface while maintaining a constant direction of analysis in the reciprocal space. The sample was first aligned using the substrate peaks (black curve in Fig.~\ref{fig.Xrays}(b)). Due to the poor intensity of the collected signal, the monochromator is not used in this configuration and different wavelengths are present in the incident radiation (Cu K$_\alpha$ and K$_\beta$, W L$_\alpha$). Such wavelengths give rise to the diffraction peaks for (001)Si identified with asterisks in the figure. The remaining peaks, of lower intensity and larger width, can only be associated to the superlattice. To confirm this, the experiment was repeated with the substrate out of the Bragg condition and we found the diffraction peaks observed previously (green curve in Fig.~\ref{fig.Xrays}(b)). Furthermore, these peaks maintained the same intensity, which means that the thin film does not have any particular crystalline texture in the plane. The superlattice peak positions are compatible with the fcc crystalline structure of Pt, and the estimated interplanar distances considering Cu K$_\alpha$ radiation wavelength are very close to those tabulated for Pt. \cite{Pt_chart, Co_chart} We also notice that the (200) orientation is virtually absent, which indicates that this orientation is particularly unfavorable in the plane.

\begin{figure}[ht]
    \includegraphics[width=0.5\columnwidth]{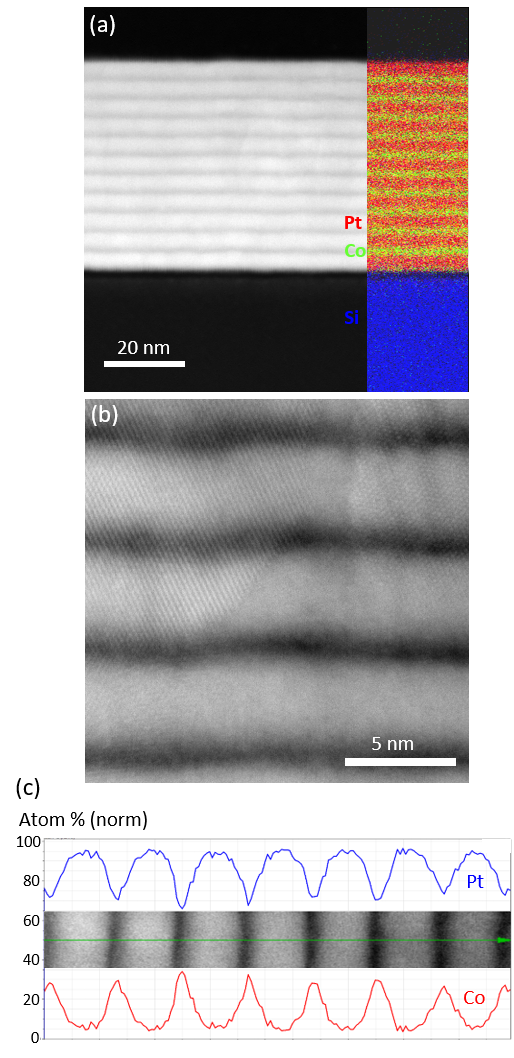}
    \caption{\footnotesize (a) Cross section view of SL4 obtained by STEM in the HAADF mode; on the right: the composition map indicates the presence of Si (substrate), Pt and Co, detected by EDS analysis. (b) High resolution HAADF-STEM image showing crystalline grains with different orientations, one grain spans along several periods of the SL. (c) Spatial distribution of Pt and Co atoms across the SL, determined by local EDS analysis; the scan-line is shown in the HAADF image in the inset.}
    \label{fig.STEM}
\end{figure}

Figure \ref{fig.STEM}(a) shows a scanning transmission electron microscopy (STEM) cross-section image of SL4 obtained with a FEI/ThermoFisher Themis Titan operating at 200 keV and equipped with a Cs spherical aberration corrector allowing atomic resolution in STEM mode and a Bruker Super-X detector for energy dispersive X-Ray spectroscopy (EDS). The cross-section has been realized by focused ion beam (FIB) using a FEI/Thermofisher dual-beam. In the high angle annular dark field (HAADF) mode, the contrast is directly correlated to the Z-number of the atoms, the brighter the heavier. The ten Pt/Co periods and the Pt capping layer are clearly distinguished. EDS analysis was used to quantify the atomic composition across the SL. The layers are continuous and homogeneous in thickness as well as in composition.
There are regions of the sample where the layers display a higher roughness, which can be associated to the effect of crystalline grains. High-resolution TEM images give account of a polycrystalline structure, with grains that may span multiple SL periods (see Fig. \ref{fig.STEM}(b)). Figure \ref{fig.STEM}(c) presents results of a local EDS chemical analysis performed by sweeping a line profile across the SL. Both Co and Pt atomic species are present along the whole sample, with a composition modulation that reaches a maximum Pt concentration of 95 at.$\%$ in Pt rich regions, and a minimum of 65 at.$\%$ in the very thin Co-rich layers. We have shown here that the SLs studied in this work present a well-defined periodic structure with interfaces characterized by a smooth contrast gradient instead of a sharp change in contrast. It is noteworthy that the TEM lamella is several tens of nm thick, which means that the apparent gradients at the interface could be due to interfacial roughness in the depth of the TEM lamella.

\section{Magnetic properties} \label{magnetic}

\begin{figure}[h]
    \centering
    \includegraphics[width=0.7\columnwidth]{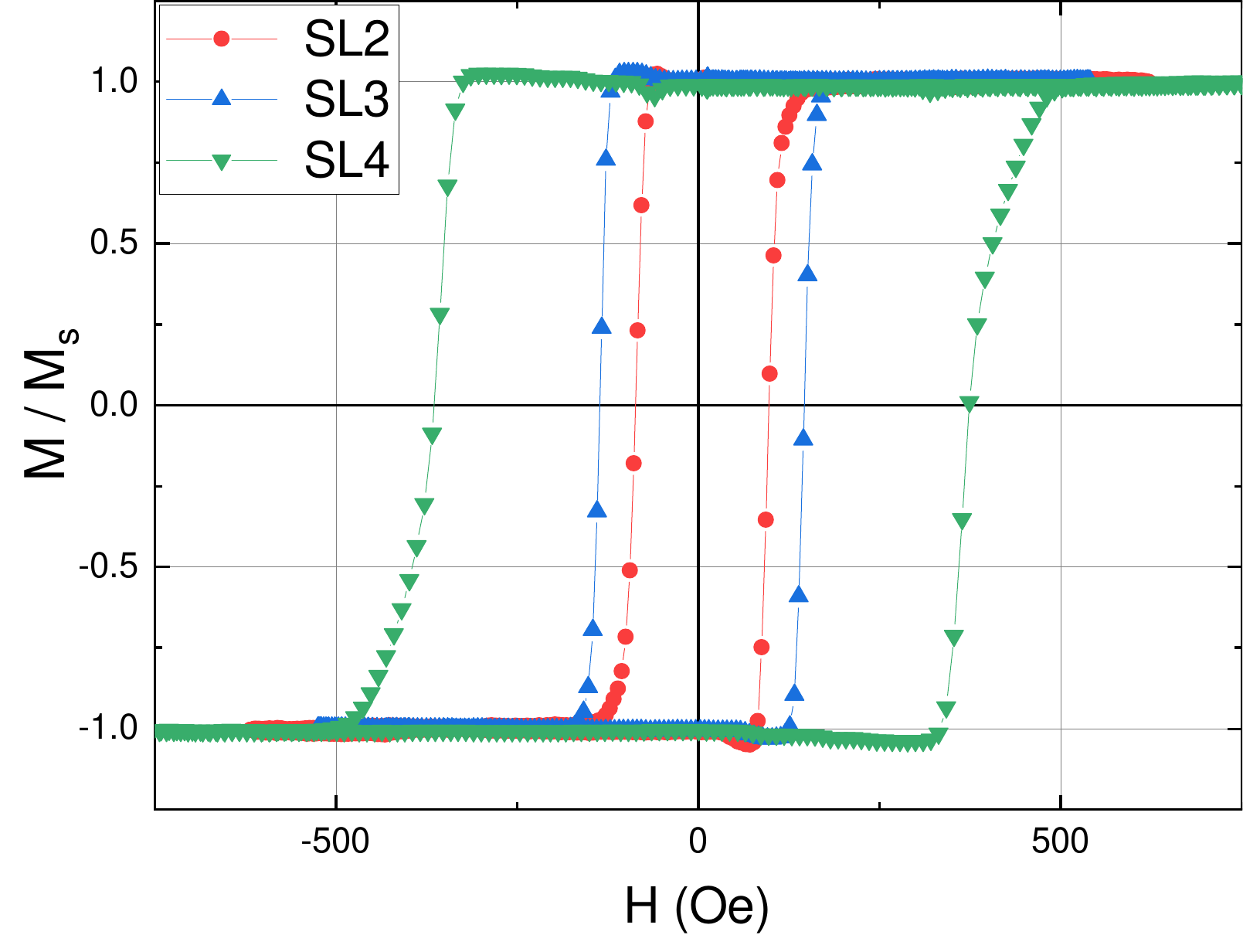}
    \caption{\footnotesize Normalized out-of-plane magnetization loops measured by Kerr magnetometry in polar configuration for SLs with different Co thicknesses. Square loops are observed in samples with $t_\text{Co}\geq 0.45$ nm.}
    \label{fig.MvsH}
\end{figure}

Magnetization loops with the magnetic field perpendicular to the plane were measured by Kerr magnetometry in polar configuration. Figure \ref{fig.MvsH} shows the normalized magnetization curves for those SLs presenting a perpendicular contribution to the magnetization. Square magnetization loops were obtained for samples with $t_\text{Co}\geq 0.45$ nm, indicating the presence of strong perpendicular magnetic anisotropy.
We observe that the coercive field $H_C$ increases with increasing Co thickness, as previously observed in Pt/Co/Pt stacks,\cite{QuinterosAPL,QuinterosGrowth} indicating an increase of the perpendicular magnetic anisotropy with $t_\text{Co}$.
The sample with $t_\text{Co} = 0.18$ nm did not present any detectable magnetization.

The saturation magnetization was quantified by SQUID magnetometry for SL4. We obtained $M_S = (1110 \pm 170)$ emu/cm$^3$ at room temperature, estimated as the ratio of the measured magnetic moment to the nominal cobalt volume in the sample. This value is slightly smaller than that for bulk cobalt, $M_S = 1420 $ emu/cm$^3$, which seems reasonable considering the poor Co concentration in the Co-rich layers described in Sec.\ref{structural}.

\section{Ultrahigh frequency acoustic phonon dynamics} \label{Phonons}

\begin{figure}[h!]
    \includegraphics[width=0.7\columnwidth]{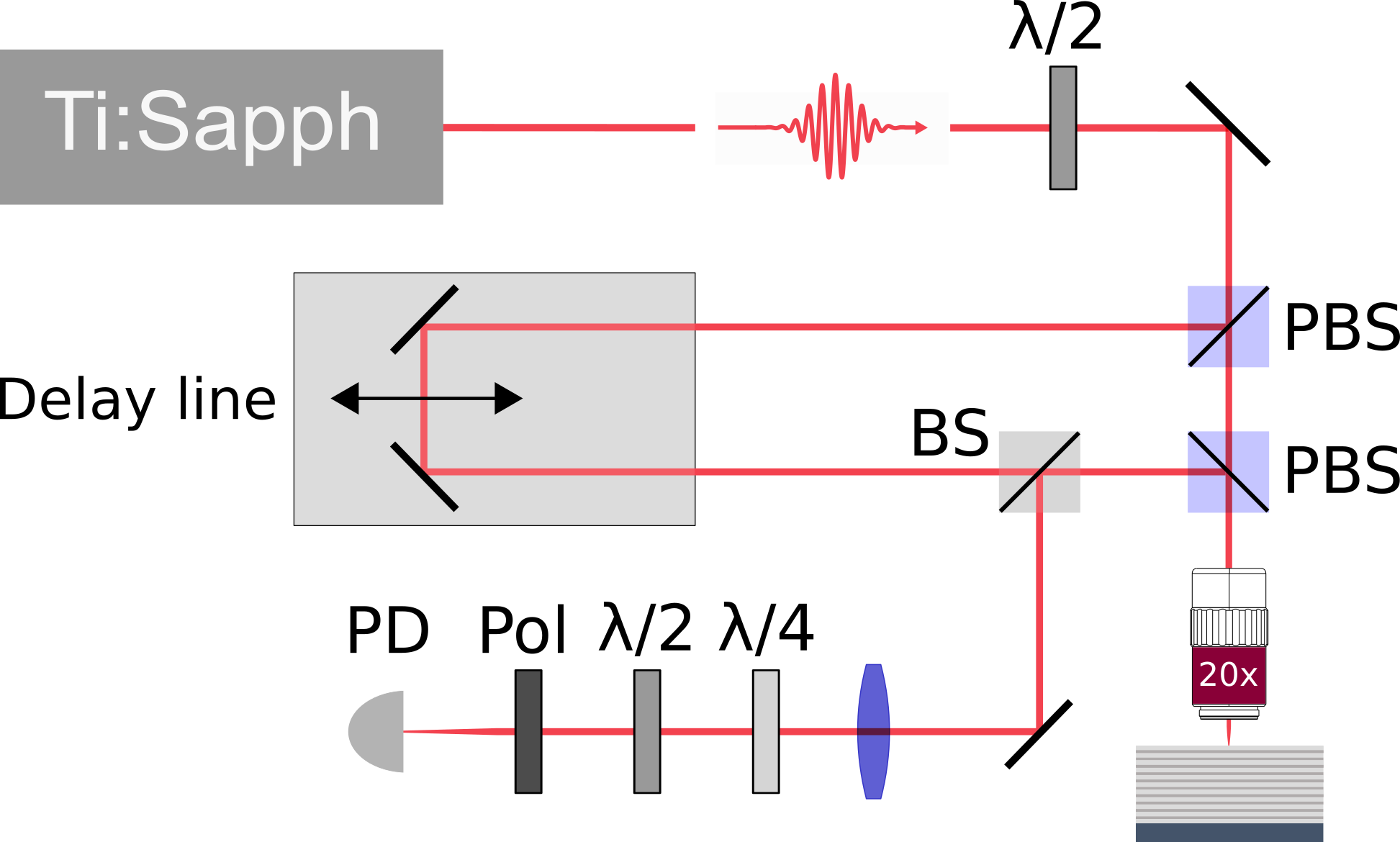}
    \caption{\footnotesize Schematics of the pump-probe experimental setup. BS, beam splitter; PBS, polarizing beam splitter; Pol, polarizer; $\lambda$/2, half-wave plate; $\lambda$/4, quarter-wave plate; PD, photodiode.}
    \label{setup}
\end{figure}

We employed a pump-probe setup (see Fig.~\ref{setup}) to experimentally study the coherent acoustic phonon dynamics in the metallic superlattices.\cite{rossignolPicosecondUltrasonicsStudy2001,thomsenSurfaceGenerationDetection1986} A Ti:sapphire laser produces $\sim150$~fs pulses at a repetition rate of 80~MHz and wavelength of 850~nm. A polarization beam splitter (PBS) divides the beam into two orthogonally polarized parts, namely the pump and the probe, following different paths. Both beams are focused on the sample through a 20X objective, resulting in a spot size of $\approx5$-$\mu$m-diameter in a normal incidence configuration. The pump pulse first impinges onto the sample and is absorbed, inducing a temperature increase in the sample. 
This temperature increase results in an ultrafast expansion, which generates a strain pulse propagating within the structure and folded acoustic phonons at the Brillouin zone center. The presence of both modes modifies the complex refractive index in the multilayer, resulting in a variation of the reflectivity ($\Delta$R) that is detected by a probe pulse arriving on the sample with a time delay.
To increase the signal-to-noise ratio, the pump is modulated at 800 kHz by an acousto-optical modulator connected with a lock-in amplifier for synchronous detection. In the collection path, the pump beam is filtered out by a combination of a half and a quarter waveplates, and a polarizer.\cite{ortizFiberintegratedMicrocavitiesEfficient2020}

\begin{figure}[hb]
    \includegraphics[width=0.7\columnwidth]{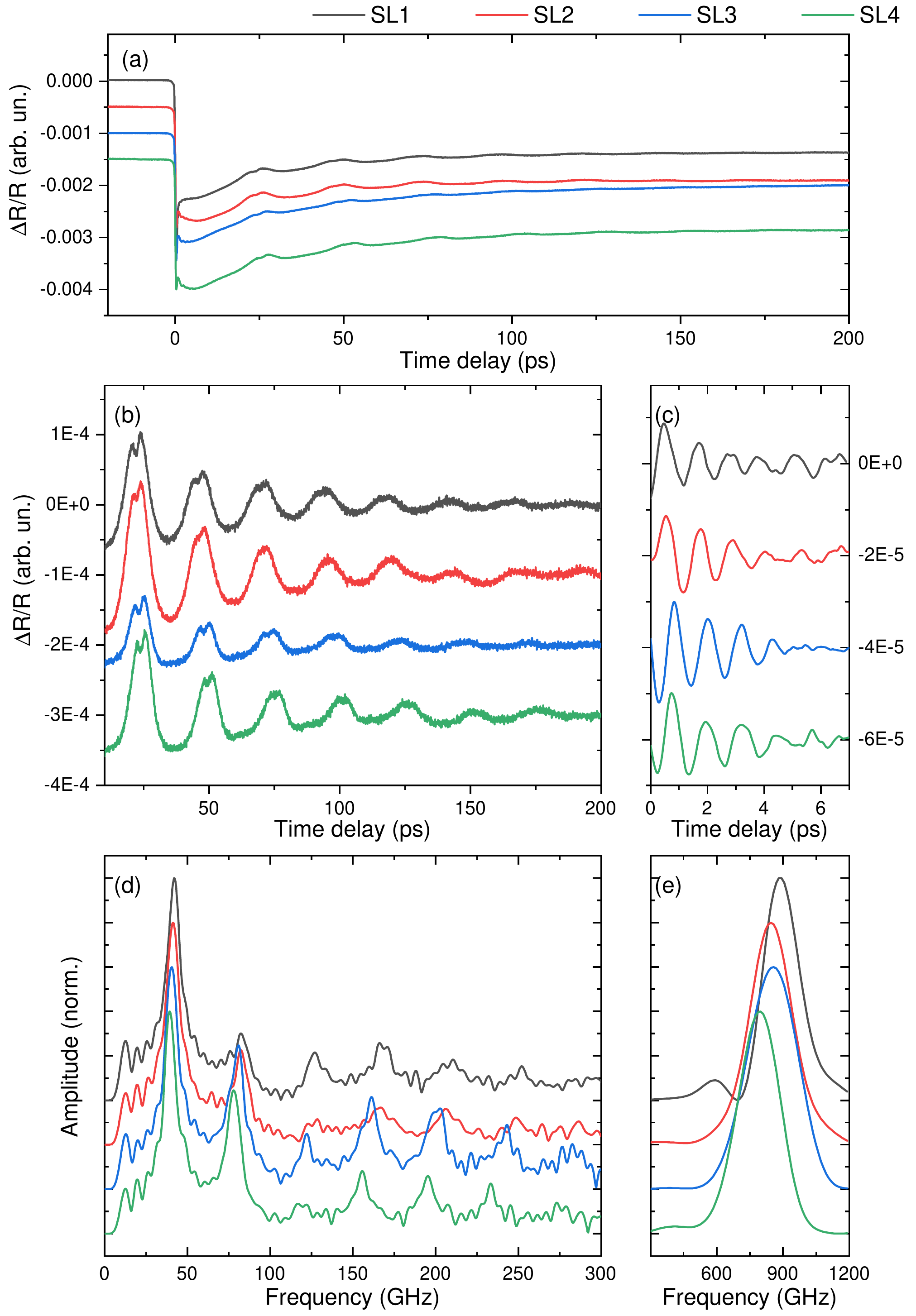}
    \caption{\footnotesize (a) Raw timetraces of SL1 (black), SL2 (red), SL3 (blue) and SL4 (green). (b) Polynomial fit between 10 and 200 ps of the timetraces displayed in (a), and (c) Band-pass filter between 500 and 1500 GHz, for the first 7 ps. (d) FFT of the timetraces displayed in (b). (e) Same as (d), but with the timetraces displayed in (c).} 
    \label{fig:timetraces}
\end{figure}

Figure~\ref{fig:timetraces}(a) shows transient reflectivity signals measured on the four superlattices. A rapid change at t~=~0~ps in $\Delta$R is caused by the pump excitation, followed by a slow relaxation of the system back to the equilibrium state. To analyze the lower- and higher-frequency components present in the timetrace, we separate the raw data from 10 to 200~ps (Fig~\ref{fig:timetraces}(b)) and from 0 to 7~ps (Fig~\ref{fig:timetraces}(c)), respectively. In the 10–200~ps region, a polynomial fit is applied to the raw data to remove the slow relaxation component, revealing oscillations with a period of approximately 25~ps, which decay in amplitude over time. By performing Fast Fourier transform, the phononic spectra displayed in Fig.~\ref{fig:timetraces}(d) are obtained, featuring a signal at $\sim$40~GHz corresponding to the fundamental mode of the whole multilayer stack as an effective single resonator. 
In addition harmonic peaks up to 250~GHz are also present.  For the first 7~ps, the short-period oscillations with rapid decay are extracted using the derivative of the raw data (Fig.\ref{fig:timetraces}(c)). The corresponding Fourier transform, presented in Fig.\ref{fig:timetraces}(e), identifies a high-frequency signal at $\sim$800~GHz.

\begin{figure}[h!]
    \includegraphics[width=0.7\columnwidth]{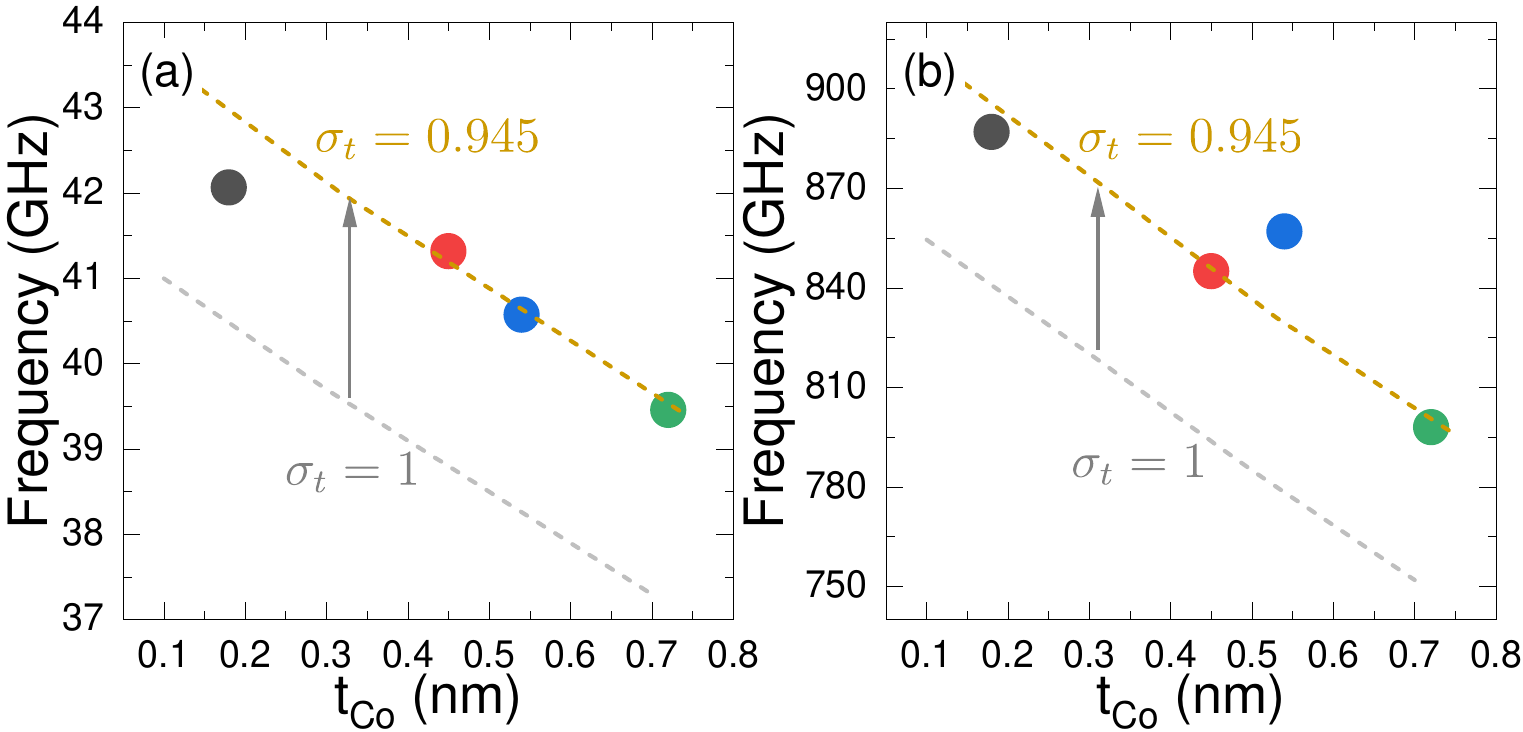}
    \caption{\footnotesize Peak position of all the superlattices for the mode at (a) $\sim 40$~GHz and (b) $\sim 800$~GHz. The dashed grey and dark yellow lines represent theoretical frequencies obtained using different correction factors $\sigma_t$ to the nominal thicknesses of Pt and Co used for the transfer matrix method simulations.}
    \label{fig:peak_position}
\end{figure}

The peak positions in the phononic spectra are extracted for the four samples. As the Co layer nominal thickness increases from 0.18 to 0.72~nm, the total sample thickness also increases, resulting in a red shift of the mode at $\sim 40$~GHz (Figs.~\ref{fig:peak_position}(a)). Similarly, the thickness of the unit cell constituting the superlattice increases, leading to a red shift of the mode at $\sim 800$~GHz as shown in Figs.~\ref{fig:peak_position}(b).

\begin{figure}[t]
    \includegraphics[width=0.7\columnwidth]{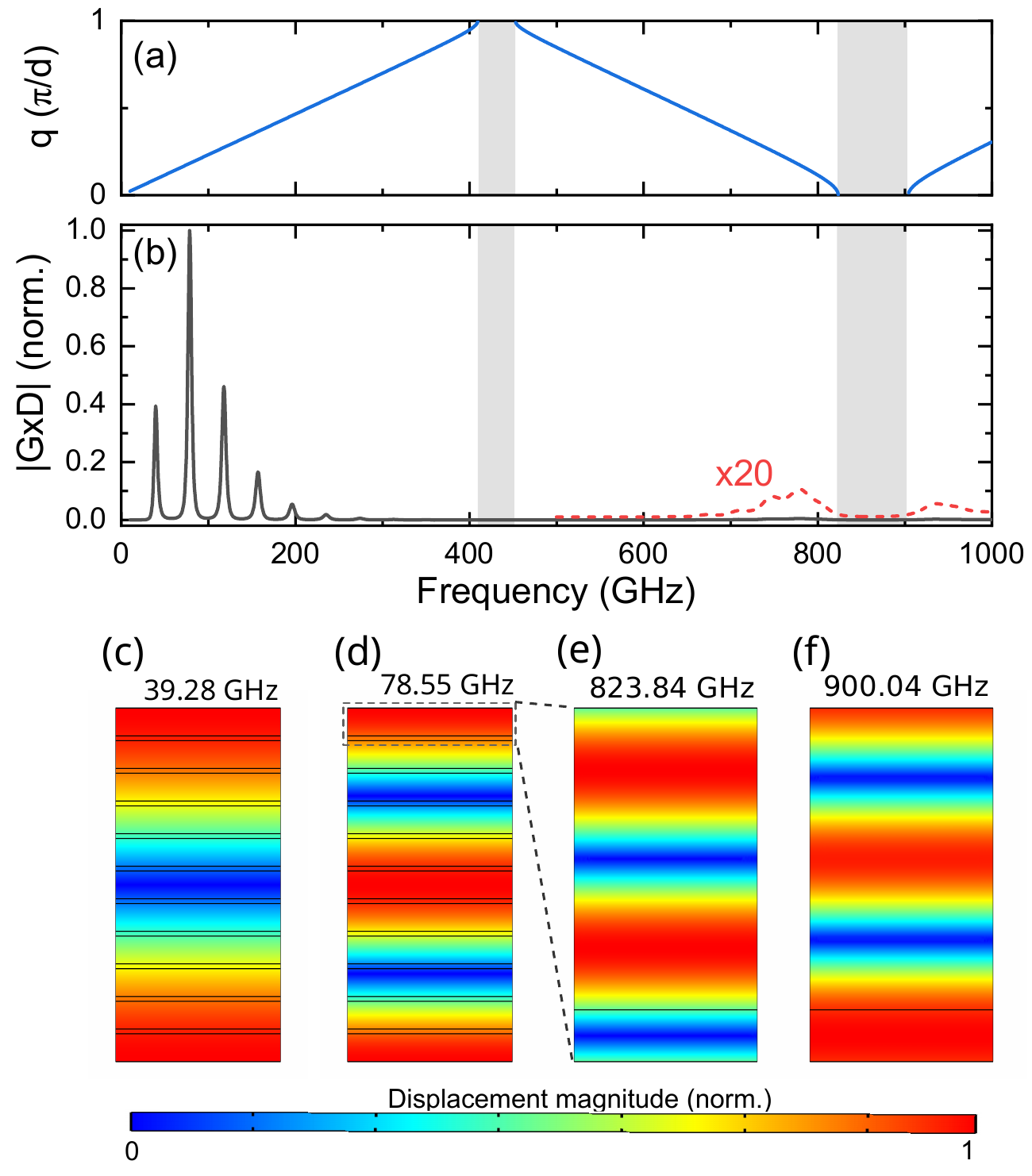}
    \caption{\footnotesize (a) Calculated acoustic dispersion relation of SL4. The stop-bands at $\sim$400 and $\sim$850 GHz are highlighted in grey. (b) Simulation of the pump-probe generation-detection spectrum. Lower frequencies are related to the full stack, higher frequencies around 800~GHz correspond to acoustic modes of the superlattice. The red-dashed line is multiplied by 20 for clarity. (c) and (d) are the displacement profile along the structure for the fundamental mode at $\sim$39 GHz and first harmonic at $\sim$79~GHz, respectively. (e) and (f) correspond to the displacement profile of the superlattice modes at $\sim$800~GHz.}
    \label{fig:sim_freq}
\end{figure}

Frequency- and time-domain simulations of the acoustic properties of SL4 were performed using both transfer matrix method, incorporating the photoelastic model, and finite element method using COMSOL.\cite{lanzillotti-kimuraTheoryCoherentGeneration2011,pascual-winterSpectralResponsesPhonon2012} Figure~\ref{fig:sim_freq}(a) illustrates the first Brillouin zone of the acoustic dispersion relation for the longitudinal mode of the superlattice SL4. Two stop-band regions are clearly observed, one near $\sim$420 GHz (zone edge) and the other at $\sim$850 GHz (zone center). The amplitude and position of the stop-band openings depend on the material properties (sound velocity and mass density) as well as on the period of the SL.

Using transfer matrix method, we calculate the pump-probe generation-detection spectrum, displayed in Figure~\ref{fig:sim_freq}(b). The details of the simulation can be found in Ref.~\citenum{lanzillotti-kimuraTheoryCoherentGeneration2011}, and the parameters used are shown in Table~\ref{table:parameters}. The spectrum reveals distinct features in two frequency ranges, between 40 GHz and 300 GHz, and above 700 GHz. 

The lower frequency peaks correspond to acoustic modes related to the total thickness of the superlattice, which acts as an acoustic resonator with a fundamental frequency of approximately 40~GHz. Figures~\ref{fig:sim_freq}(c) and \ref{fig:sim_freq}(d) show the displacement profiles of the fundamental mode (39.28~GHz) and its first harmonic (78.55~GHz), corresponding to the first and second peaks of panel~\ref{fig:sim_freq}(b). In both cases, the displacement profiles are commensurate with the total thickness of the SL.

In contrast, the higher frequency peaks are linked to the lower and upper band superlattice modes at the Brillouin zone center. These modes are displayed in panels~\ref{fig:sim_freq}(e) and \ref{fig:sim_freq}(f), where the displacement profiles are commensurate with the period of the SL, as well as with the total thickness.

\begin{table}
	\caption{\footnotesize Parameters used in the simulations. $n$, $v$, $\rho$ and $\delta$ corresponds to refractive index, sound velocity, mass density and penetration depth. The longitudinal sound velocity of Pt and Co are calculated considering the Young modulus ($E$), Poisson ratio ($\nu$) and mass density ($\rho$), obtained from the database of COMSOL and reference~[\citenum{CobaltCo} ], respectively, according to $v=\sqrt{\frac{E(1-\nu)}{\rho(1+\nu)(1-2\nu)}}$ .}
	\resizebox{0.7\columnwidth}{!}{\begin{tabular}{|c|c|c|c|c|}
	
		\hline
		Material & $n$ & $v$ (m/s) & $\rho$ (g/cm$^{3}$) & $\delta$ (nm) \\
		\hline
		Pt & 3.01 + 5.19i\cite{rakicOpticalPropertiesMetallic1998,polyanskiyRefractiveindexinfoDatabaseOptical2024} & 3829.06 & 21.45 & 150 \\
		\hline
		Co & 2.57 + 4.99i\cite{johnsonOpticalConstantsTransition1974,polyanskiyRefractiveindexinfoDatabaseOptical2024} & 5857.55 & 8.8 & 150 \\
		\hline
		Si\cite{cardozodeoliveiraProbingGigahertzCoherent2023} & 3.72 & 8433 & 2.32 & inf \\
		\hline
	\end{tabular}}
	\label{table:parameters}
\end{table}

We also simulated the frequency dependence of the fundamental mode and the superlattice mode on the cobalt thickness $t_\text{Co}$. The results, represented by dashed lines, are shown in Figs.~\ref{fig:peak_position}(a) and \ref{fig:peak_position}(b), respectively. In order to match the simulation results with the experimental data, we introduce a correction factor with respect to the nominal thickness ($\sigma_{t}$). Using the nominal Pt and Co thicknesses ($\sigma_{t}=1$) leads to a deviation of approximately 5.5~\% between the experimental data and the theoretical results. Applying a correction factor of $\sigma_{t}=0.945$ significantly improves agreement with experimental results in both the lower and higher frequency ranges. The differences might stem from different factors, such as deviations in sound velocities for thin films, interdiffusion of Pt and Co atoms, and material roughness, which are not taken into account in the simulations (though clearly present in the studied SLs, as reported in Sec.\ref{structural}), and could lead to position-dependent frequencies. Nevertheless, the consistency of the results still validates the underlying simulation approach and confirms its reliability in capturing the essential physical phenomena.

\begin{figure}[h!]
    \includegraphics[width=0.7\columnwidth]{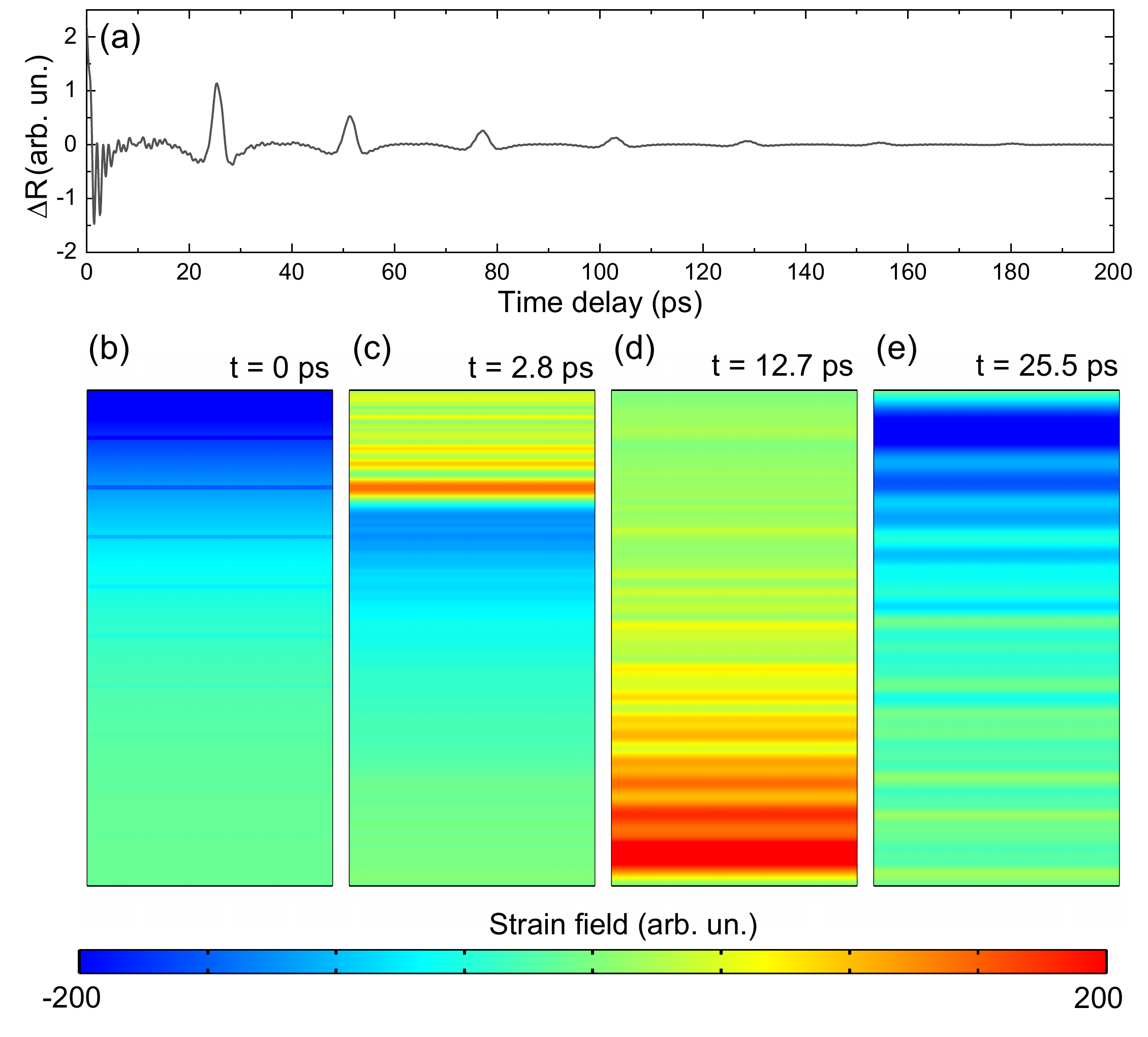}
    \caption{\footnotesize (a) Simulated timetrace using transfer matrix method. (b)-(e) Strain field along the structure at (b) $t=0$~ps (c) $t=2.8$~ps (d) $t=12.7$~ps and (e) $t=25.5$~ps, simulated with COMSOL.}
    \label{fig:sim_time}
\end{figure}

The simulations in the time domain were conducted using two approaches. The first approach involves simulating an initial strain field containing spectral components from the generation spectrum, calculated over the 0-1000 GHz frequency range. By applying an inverse Fourier transform, the temporal evolution of the strain field along the structure is obtained. Subsequently, we compute the initial optical reflectivity, $r_0$, and the perturbed reflectivity, $r'$. For simplicity, the reflectivity is calculated for a single wavelength, $\lambda = 850$~nm, which corresponds to the central laser wavelength used in the experiment for the excitation and detection beams. The initial reflectivity, $r_0$, represents the reflectivity of the unperturbed structure, i.e., without any strain-induced modifications. To compute $r'$, the strain field at each time step is used to perturb the refractive index along the structure. This perturbation is modeled as $n' = n_0 + \Delta n$, where $\Delta n \propto \frac{\mathrm{strain}}{n_0}$. Finally, the change in reflectivity is determined as $\Delta R = |r_0|^2 - |r'|^2$. The simulation was performed up to 200~ps. The results are shown in Fig.~\ref{fig:sim_time}(a). It is worth noting the similarities with the experimental time trace displayed in Figs.~\ref{fig:timetraces}(a)-(c), in particular the fast decay of the high-frequency oscillations in the beginning of the time trace, and the periodic features every $\sim$25~ps.

In the second approach, performed with COMSOL, it is imposed as initial condition a strain field that is maximum at the surface, followed by a gradual decay through the structure, to emulate the optical skin depth of the structure. The skin depth is estimated by simulating the electric field of the laser, considering the complex refractive index of Pt and Co, displayed in Table~\ref{table:parameters}. The time evolution of the strain field is then computed. Panels~\ref{fig:sim_time}(b)-(d) display the result of the strain field at (b) $t=0$~ps (c) $t=2.8$~ps (d) $t=12.7$~ps and (e) $t=25.5$~ps.

Panel ~\ref{fig:sim_time}(c) shows the strain propagation through the material, unveiling information of the superlattice modes. Panel~\ref{fig:sim_time}(d) depicts the strain field reaching the bottom of the superlattice. Finally, panel~\ref{fig:sim_time}(e) shows the acoustic strain reflected back towards the surface, where the detection efficiency is higher. The round trip of the strain pulse corresponds to the periodic modulation observed in both experimental and simulated timetraces.

\section{Conclusion and perspectives} \label{conclusion}

In summary, we have investigated the structural, magnetic, and acoustic properties of four Pt/Co-based ferromagnetic superlattices with varying cobalt thicknesses. X-ray reflectometry and STEM imaging confirmed the presence of sub-5~nm-thick periodic layers of Pt and Co with high structural quality. These superlattices exhibited clear magnetization hysteresis behavior, with properties strongly dependent on the Co layer thickness. In this multilayered structure, we have demonstrated the generation and detection of ultrahigh-frequency acoustic phonons close to 900~GHz, along with complex phonon dynamics at frequencies below 300~GHz. Simulations based on transfer matrix method and finite element method have been performed to characterize the acoustic modes and their dynamics, further supporting our experimental results. The cobalt thickness dependence of phonon modes highlights the ability of these structures for phonon engineering at the nanoscale, compatible with the stringent nanophononic requirements.

Significant acoustic absorption in Co/Pt superlattices remains challenging for extending oscillation cycles of the acoustic response.
To address this, future work could explore substrates with higher acoustic impedance contrasts to mitigate phonon leakage, as well as optimize deposition techniques to enhance interface quality. Additionally, investigating laser polarization as an extra degree of freedom might be an asset on controlling acoustic phonon generation and propagation in ferromagnetic superlattices.~\cite{lanzillotti-kimuraPolarizationcontrolledCoherentPhonon2018} Furthermore, such nanoacoustic systems can be potentially integrated with optical fibers, offering enhanced stability and reproducibility of experiments.~\cite{ortizFiberintegratedMicrocavitiesEfficient2020} Despite current limitations, our findings represent significant progress toward the integration of ultrahigh-frequency acoustic phonons and magnons for magnophononic/magnonanomechanics applications, highlighting the potential of nanoscale superlattices for controlling magnetic properties using acoustic fields.







\section {acknowledgement}
The authors acknowledge support from the C.N.R.S. International Research Project Phenomenas. E.R.C. de O., C.X., S.S., and N.D.L.-K. acknowledge funding from European Research Council Consolidator Grant No.101045089 (T-Recs). M.G. and J.E.G acknowledge support from MSCA-RISE-H2020 ULTIMATE-I-Project No 101007825 funded by the European Union. 
The authors acknowledge the French RENATECH network which partly funds the XRD characterizations, FIB
preparation, TEM observations carried out in this work.
\bibliography{references}

\begin{thebibliography}{56}%
\makeatletter
\providecommand \@ifxundefined [1]{%
 \@ifx{#1\undefined}
}%
\providecommand \@ifnum [1]{%
 \ifnum #1\expandafter \@firstoftwo
 \else \expandafter \@secondoftwo
 \fi
}%
\providecommand \@ifx [1]{%
 \ifx #1\expandafter \@firstoftwo
 \else \expandafter \@secondoftwo
 \fi
}%
\providecommand \natexlab [1]{#1}%
\providecommand \enquote  [1]{``#1''}%
\providecommand \bibnamefont  [1]{#1}%
\providecommand \bibfnamefont [1]{#1}%
\providecommand \citenamefont [1]{#1}%
\providecommand \href@noop [0]{\@secondoftwo}%
\providecommand \href [0]{\begingroup \@sanitize@url \@href}%
\providecommand \@href[1]{\@@startlink{#1}\@@href}%
\providecommand \@@href[1]{\endgroup#1\@@endlink}%
\providecommand \@sanitize@url [0]{\catcode `\\12\catcode `\$12\catcode `\&12\catcode `\#12\catcode `\^12\catcode `\_12\catcode `\%12\relax}%
\providecommand \@@startlink[1]{}%
\providecommand \@@endlink[0]{}%
\providecommand \url  [0]{\begingroup\@sanitize@url \@url }%
\providecommand \@url [1]{\endgroup\@href {#1}{\urlprefix }}%
\providecommand \urlprefix  [0]{URL }%
\providecommand \Eprint [0]{\href }%
\providecommand \doibase [0]{https://doi.org/}%
\providecommand \selectlanguage [0]{\@gobble}%
\providecommand \bibinfo  [0]{\@secondoftwo}%
\providecommand \bibfield  [0]{\@secondoftwo}%
\providecommand \translation [1]{[#1]}%
\providecommand \BibitemOpen [0]{}%
\providecommand \bibitemStop [0]{}%
\providecommand \bibitemNoStop [0]{.\EOS\space}%
\providecommand \EOS [0]{\spacefactor3000\relax}%
\providecommand \BibitemShut  [1]{\csname bibitem#1\endcsname}%
\let\auto@bib@innerbib\@empty
\bibitem [{\citenamefont {Comstock}(2002)}]{comstockReviewModernMagnetic2002}%
  \BibitemOpen
  \bibfield  {author} {\bibinfo {author} {\bibfnamefont {R.~L.}\ \bibnamefont {Comstock}},\ }\bibfield  {title} {\enquote {\bibinfo {title} {Review modern magnetic materials in data storage},}\ }\href {https://doi.org/10.1023/A:1019642215245} {\bibfield  {journal} {\bibinfo  {journal} {Journal of Materials Science: Materials in Electronics}\ }\textbf {\bibinfo {volume} {13}},\ \bibinfo {pages} {509--523} (\bibinfo {year} {2002})}\BibitemShut {NoStop}%
\bibitem [{\citenamefont {Beaurepaire}\ \emph {et~al.}(1996)\citenamefont {Beaurepaire}, \citenamefont {Merle}, \citenamefont {Daunois},\ and\ \citenamefont {Bigot}}]{beaurepaireUltrafastSpinDynamics1996}%
  \BibitemOpen
  \bibfield  {author} {\bibinfo {author} {\bibfnamefont {E.}~\bibnamefont {Beaurepaire}}, \bibinfo {author} {\bibfnamefont {J.-C.}\ \bibnamefont {Merle}}, \bibinfo {author} {\bibfnamefont {A.}~\bibnamefont {Daunois}},\ and\ \bibinfo {author} {\bibfnamefont {J.-Y.}\ \bibnamefont {Bigot}},\ }\bibfield  {title} {\enquote {\bibinfo {title} {Ultrafast spin dynamics in ferromagnetic nickel},}\ }\href {https://doi.org/10.1103/PhysRevLett.76.4250} {\bibfield  {journal} {\bibinfo  {journal} {Physical Review Letters}\ }\textbf {\bibinfo {volume} {76}},\ \bibinfo {pages} {4250--4253} (\bibinfo {year} {1996})}\BibitemShut {NoStop}%
\bibitem [{\citenamefont {Bigot}\ \emph {et~al.}(2004)\citenamefont {Bigot}, \citenamefont {Guidoni}, \citenamefont {Beaurepaire},\ and\ \citenamefont {Saeta}}]{bigotFemtosecondSpectrotemporalMagnetooptics2004}%
  \BibitemOpen
  \bibfield  {author} {\bibinfo {author} {\bibfnamefont {J.-Y.}\ \bibnamefont {Bigot}}, \bibinfo {author} {\bibfnamefont {L.}~\bibnamefont {Guidoni}}, \bibinfo {author} {\bibfnamefont {E.}~\bibnamefont {Beaurepaire}},\ and\ \bibinfo {author} {\bibfnamefont {P.~N.}\ \bibnamefont {Saeta}},\ }\bibfield  {title} {\enquote {\bibinfo {title} {Femtosecond spectrotemporal magneto-optics},}\ }\href {https://doi.org/10.1103/PhysRevLett.93.077401} {\bibfield  {journal} {\bibinfo  {journal} {Physical Review Letters}\ }\textbf {\bibinfo {volume} {93}},\ \bibinfo {pages} {077401} (\bibinfo {year} {2004})}\BibitemShut {NoStop}%
\bibitem [{\citenamefont {Weisheit}\ \emph {et~al.}(2007)\citenamefont {Weisheit}, \citenamefont {F{\"a}hler}, \citenamefont {Marty}, \citenamefont {Souche}, \citenamefont {Poinsignon},\ and\ \citenamefont {Givord}}]{weisheitElectricFieldInducedModification2007}%
  \BibitemOpen
  \bibfield  {author} {\bibinfo {author} {\bibfnamefont {M.}~\bibnamefont {Weisheit}}, \bibinfo {author} {\bibfnamefont {S.}~\bibnamefont {F{\"a}hler}}, \bibinfo {author} {\bibfnamefont {A.}~\bibnamefont {Marty}}, \bibinfo {author} {\bibfnamefont {Y.}~\bibnamefont {Souche}}, \bibinfo {author} {\bibfnamefont {C.}~\bibnamefont {Poinsignon}},\ and\ \bibinfo {author} {\bibfnamefont {D.}~\bibnamefont {Givord}},\ }\bibfield  {title} {\enquote {\bibinfo {title} {Electric field-induced modification of magnetism in thin-film ferromagnets},}\ }\href {https://doi.org/10.1126/science.1136629} {\bibfield  {journal} {\bibinfo  {journal} {Science}\ }\textbf {\bibinfo {volume} {315}},\ \bibinfo {pages} {349--351} (\bibinfo {year} {2007})}\BibitemShut {NoStop}%
\bibitem [{\citenamefont {Chu}\ \emph {et~al.}(2008)\citenamefont {Chu}, \citenamefont {Martin}, \citenamefont {Holcomb}, \citenamefont {Gajek}, \citenamefont {Han}, \citenamefont {He}, \citenamefont {Balke}, \citenamefont {Yang}, \citenamefont {Lee}, \citenamefont {Hu}, \citenamefont {Zhan}, \citenamefont {Yang}, \citenamefont {{Fraile-Rodr{\'i}guez}}, \citenamefont {Scholl}, \citenamefont {Wang},\ and\ \citenamefont {Ramesh}}]{chuElectricfieldControlLocal2008}%
  \BibitemOpen
  \bibfield  {author} {\bibinfo {author} {\bibfnamefont {Y.-H.}\ \bibnamefont {Chu}}, \bibinfo {author} {\bibfnamefont {L.~W.}\ \bibnamefont {Martin}}, \bibinfo {author} {\bibfnamefont {M.~B.}\ \bibnamefont {Holcomb}}, \bibinfo {author} {\bibfnamefont {M.}~\bibnamefont {Gajek}}, \bibinfo {author} {\bibfnamefont {S.-J.}\ \bibnamefont {Han}}, \bibinfo {author} {\bibfnamefont {Q.}~\bibnamefont {He}}, \bibinfo {author} {\bibfnamefont {N.}~\bibnamefont {Balke}}, \bibinfo {author} {\bibfnamefont {C.-H.}\ \bibnamefont {Yang}}, \bibinfo {author} {\bibfnamefont {D.}~\bibnamefont {Lee}}, \bibinfo {author} {\bibfnamefont {W.}~\bibnamefont {Hu}}, \bibinfo {author} {\bibfnamefont {Q.}~\bibnamefont {Zhan}}, \bibinfo {author} {\bibfnamefont {P.-L.}\ \bibnamefont {Yang}}, \bibinfo {author} {\bibfnamefont {A.}~\bibnamefont {{Fraile-Rodr{\'i}guez}}}, \bibinfo {author} {\bibfnamefont {A.}~\bibnamefont {Scholl}}, \bibinfo {author} {\bibfnamefont {S.~X.}\ \bibnamefont {Wang}},\ and\ \bibinfo {author} {\bibfnamefont
  {R.}~\bibnamefont {Ramesh}},\ }\bibfield  {title} {\enquote {\bibinfo {title} {Electric-field control of local ferromagnetism using a magnetoelectric multiferroic},}\ }\href {https://doi.org/10.1038/nmat2184} {\bibfield  {journal} {\bibinfo  {journal} {Nature Materials}\ }\textbf {\bibinfo {volume} {7}},\ \bibinfo {pages} {478--482} (\bibinfo {year} {2008})}\BibitemShut {NoStop}%
\bibitem [{\citenamefont {Rana}\ and\ \citenamefont {Otani}(2019)}]{ranaMagnonicDevicesBased2019}%
  \BibitemOpen
  \bibfield  {author} {\bibinfo {author} {\bibfnamefont {B.}~\bibnamefont {Rana}}\ and\ \bibinfo {author} {\bibfnamefont {Y.}~\bibnamefont {Otani}},\ }\bibfield  {title} {\enquote {\bibinfo {title} {Towards magnonic devices based on voltage-controlled magnetic anisotropy},}\ }\href {https://doi.org/10.1038/s42005-019-0189-6} {\bibfield  {journal} {\bibinfo  {journal} {Communications Physics}\ }\textbf {\bibinfo {volume} {2}},\ \bibinfo {pages} {1--12} (\bibinfo {year} {2019})}\BibitemShut {NoStop}%
\bibitem [{\citenamefont {Kittel}(1958)}]{kittelInteractionSpinWaves1958}%
  \BibitemOpen
  \bibfield  {author} {\bibinfo {author} {\bibfnamefont {C.}~\bibnamefont {Kittel}},\ }\bibfield  {title} {\enquote {\bibinfo {title} {Interaction of spin waves and ultrasonic waves in ferromagnetic crystals},}\ }\href {https://doi.org/10.1103/PhysRev.110.836} {\bibfield  {journal} {\bibinfo  {journal} {Physical Review}\ }\textbf {\bibinfo {volume} {110}},\ \bibinfo {pages} {836--841} (\bibinfo {year} {1958})}\BibitemShut {NoStop}%
\bibitem [{\citenamefont {Li}\ \emph {et~al.}(2021)\citenamefont {Li}, \citenamefont {Zhao}, \citenamefont {Zhang}, \citenamefont {Hoffmann},\ and\ \citenamefont {Novosad}}]{liAdvancesCoherentCoupling2021}%
  \BibitemOpen
  \bibfield  {author} {\bibinfo {author} {\bibfnamefont {Y.}~\bibnamefont {Li}}, \bibinfo {author} {\bibfnamefont {C.}~\bibnamefont {Zhao}}, \bibinfo {author} {\bibfnamefont {W.}~\bibnamefont {Zhang}}, \bibinfo {author} {\bibfnamefont {A.}~\bibnamefont {Hoffmann}},\ and\ \bibinfo {author} {\bibfnamefont {V.}~\bibnamefont {Novosad}},\ }\bibfield  {title} {\enquote {\bibinfo {title} {Advances in coherent coupling between magnons and acoustic phonons},}\ }\href {https://doi.org/10.1063/5.0047054} {\bibfield  {journal} {\bibinfo  {journal} {APL Materials}\ }\textbf {\bibinfo {volume} {9}},\ \bibinfo {pages} {060902} (\bibinfo {year} {2021})}\BibitemShut {NoStop}%
\bibitem [{\citenamefont {Yang}\ and\ \citenamefont {Schmidt}(2021)}]{yangAcousticControlMagnetism2021}%
  \BibitemOpen
  \bibfield  {author} {\bibinfo {author} {\bibfnamefont {W.-G.}\ \bibnamefont {Yang}}\ and\ \bibinfo {author} {\bibfnamefont {H.}~\bibnamefont {Schmidt}},\ }\bibfield  {title} {\enquote {\bibinfo {title} {Acoustic control of magnetism toward energy-efficient applications},}\ }\href {https://doi.org/10.1063/5.0042138} {\bibfield  {journal} {\bibinfo  {journal} {Applied Physics Reviews}\ }\textbf {\bibinfo {volume} {8}},\ \bibinfo {pages} {021304} (\bibinfo {year} {2021})}\BibitemShut {NoStop}%
\bibitem [{\citenamefont {Hioki}, \citenamefont {Hashimoto},\ and\ \citenamefont {Saitoh}(2022)}]{hiokiCoherentOscillationPhonons2022}%
  \BibitemOpen
  \bibfield  {author} {\bibinfo {author} {\bibfnamefont {T.}~\bibnamefont {Hioki}}, \bibinfo {author} {\bibfnamefont {Y.}~\bibnamefont {Hashimoto}},\ and\ \bibinfo {author} {\bibfnamefont {E.}~\bibnamefont {Saitoh}},\ }\bibfield  {title} {\enquote {\bibinfo {title} {Coherent oscillation between phonons and magnons},}\ }\href {https://doi.org/10.1038/s42005-022-00888-1} {\bibfield  {journal} {\bibinfo  {journal} {Communications Physics}\ }\textbf {\bibinfo {volume} {5}},\ \bibinfo {pages} {1--8} (\bibinfo {year} {2022})}\BibitemShut {NoStop}%
\bibitem [{\citenamefont {Weber}\ \emph {et~al.}(2022)\citenamefont {Weber}, \citenamefont {Guennou}, \citenamefont {Evans}, \citenamefont {Toulouse}, \citenamefont {Simonov}, \citenamefont {Kholina}, \citenamefont {Ma}, \citenamefont {Ren}, \citenamefont {Cao}, \citenamefont {Carpenter}, \citenamefont {Dkhil}, \citenamefont {Fiebig},\ and\ \citenamefont {Kreisel}}]{weberEmergingSpinPhonon2022}%
  \BibitemOpen
  \bibfield  {author} {\bibinfo {author} {\bibfnamefont {M.~C.}\ \bibnamefont {Weber}}, \bibinfo {author} {\bibfnamefont {M.}~\bibnamefont {Guennou}}, \bibinfo {author} {\bibfnamefont {D.~M.}\ \bibnamefont {Evans}}, \bibinfo {author} {\bibfnamefont {C.}~\bibnamefont {Toulouse}}, \bibinfo {author} {\bibfnamefont {A.}~\bibnamefont {Simonov}}, \bibinfo {author} {\bibfnamefont {Y.}~\bibnamefont {Kholina}}, \bibinfo {author} {\bibfnamefont {X.}~\bibnamefont {Ma}}, \bibinfo {author} {\bibfnamefont {W.}~\bibnamefont {Ren}}, \bibinfo {author} {\bibfnamefont {S.}~\bibnamefont {Cao}}, \bibinfo {author} {\bibfnamefont {M.~A.}\ \bibnamefont {Carpenter}}, \bibinfo {author} {\bibfnamefont {B.}~\bibnamefont {Dkhil}}, \bibinfo {author} {\bibfnamefont {M.}~\bibnamefont {Fiebig}},\ and\ \bibinfo {author} {\bibfnamefont {J.}~\bibnamefont {Kreisel}},\ }\bibfield  {title} {\enquote {\bibinfo {title} {Emerging spin--phonon coupling through cross-talk of two magnetic sublattices},}\ }\href
  {https://doi.org/10.1038/s41467-021-27267-8} {\bibfield  {journal} {\bibinfo  {journal} {Nature Communications}\ }\textbf {\bibinfo {volume} {13}},\ \bibinfo {pages} {443} (\bibinfo {year} {2022})}\BibitemShut {NoStop}%
\bibitem [{\citenamefont {Kalashnikova}\ \emph {et~al.}(2007)\citenamefont {Kalashnikova}, \citenamefont {Kimel}, \citenamefont {Pisarev}, \citenamefont {Gridnev}, \citenamefont {Kirilyuk},\ and\ \citenamefont {Rasing}}]{kalashnikovaImpulsiveGenerationCoherent2007}%
  \BibitemOpen
  \bibfield  {author} {\bibinfo {author} {\bibfnamefont {A.~M.}\ \bibnamefont {Kalashnikova}}, \bibinfo {author} {\bibfnamefont {A.~V.}\ \bibnamefont {Kimel}}, \bibinfo {author} {\bibfnamefont {R.~V.}\ \bibnamefont {Pisarev}}, \bibinfo {author} {\bibfnamefont {V.~N.}\ \bibnamefont {Gridnev}}, \bibinfo {author} {\bibfnamefont {A.}~\bibnamefont {Kirilyuk}},\ and\ \bibinfo {author} {\bibfnamefont {{\relax Th}.}~\bibnamefont {Rasing}},\ }\bibfield  {title} {\enquote {\bibinfo {title} {Impulsive generation of coherent magnons by linearly polarized light in the easy-plane antiferromagnet {{FeBO}}$_3$},}\ }\href {https://doi.org/10.1103/PhysRevLett.99.167205} {\bibfield  {journal} {\bibinfo  {journal} {Physical Review Letters}\ }\textbf {\bibinfo {volume} {99}},\ \bibinfo {pages} {167205} (\bibinfo {year} {2007})}\BibitemShut {NoStop}%
\bibitem [{\citenamefont {Scherbakov}\ \emph {et~al.}(2010)\citenamefont {Scherbakov}, \citenamefont {Salasyuk}, \citenamefont {Akimov}, \citenamefont {Liu}, \citenamefont {Bombeck}, \citenamefont {Br{\"u}ggemann}, \citenamefont {Yakovlev}, \citenamefont {Sapega}, \citenamefont {Furdyna},\ and\ \citenamefont {Bayer}}]{scherbakovCoherentMagnetizationPrecession2010}%
  \BibitemOpen
  \bibfield  {author} {\bibinfo {author} {\bibfnamefont {A.~V.}\ \bibnamefont {Scherbakov}}, \bibinfo {author} {\bibfnamefont {A.~S.}\ \bibnamefont {Salasyuk}}, \bibinfo {author} {\bibfnamefont {A.~V.}\ \bibnamefont {Akimov}}, \bibinfo {author} {\bibfnamefont {X.}~\bibnamefont {Liu}}, \bibinfo {author} {\bibfnamefont {M.}~\bibnamefont {Bombeck}}, \bibinfo {author} {\bibfnamefont {C.}~\bibnamefont {Br{\"u}ggemann}}, \bibinfo {author} {\bibfnamefont {D.~R.}\ \bibnamefont {Yakovlev}}, \bibinfo {author} {\bibfnamefont {V.~F.}\ \bibnamefont {Sapega}}, \bibinfo {author} {\bibfnamefont {J.~K.}\ \bibnamefont {Furdyna}},\ and\ \bibinfo {author} {\bibfnamefont {M.}~\bibnamefont {Bayer}},\ }\bibfield  {title} {\enquote {\bibinfo {title} {Coherent magnetization precession in ferromagnetic {{(Ga,Mn)As}} induced by picosecond acoustic pulses},}\ }\href {https://doi.org/10.1103/PhysRevLett.105.117204} {\bibfield  {journal} {\bibinfo  {journal} {Physical Review Letters}\ }\textbf {\bibinfo {volume} {105}},\ \bibinfo {pages}
  {117204} (\bibinfo {year} {2010})}\BibitemShut {NoStop}%
\bibitem [{\citenamefont {Thevenard}\ \emph {et~al.}(2010)\citenamefont {Thevenard}, \citenamefont {Peronne}, \citenamefont {Gourdon}, \citenamefont {Testelin}, \citenamefont {Cubukcu}, \citenamefont {Charron}, \citenamefont {Vincent}, \citenamefont {Lema{\^i}tre},\ and\ \citenamefont {Perrin}}]{thevenardEffectPicosecondStrain2010}%
  \BibitemOpen
  \bibfield  {author} {\bibinfo {author} {\bibfnamefont {L.}~\bibnamefont {Thevenard}}, \bibinfo {author} {\bibfnamefont {E.}~\bibnamefont {Peronne}}, \bibinfo {author} {\bibfnamefont {C.}~\bibnamefont {Gourdon}}, \bibinfo {author} {\bibfnamefont {C.}~\bibnamefont {Testelin}}, \bibinfo {author} {\bibfnamefont {M.}~\bibnamefont {Cubukcu}}, \bibinfo {author} {\bibfnamefont {E.}~\bibnamefont {Charron}}, \bibinfo {author} {\bibfnamefont {S.}~\bibnamefont {Vincent}}, \bibinfo {author} {\bibfnamefont {A.}~\bibnamefont {Lema{\^i}tre}},\ and\ \bibinfo {author} {\bibfnamefont {B.}~\bibnamefont {Perrin}},\ }\bibfield  {title} {\enquote {\bibinfo {title} {Effect of picosecond strain pulses on thin layers of the ferromagnetic semiconductor ({{Ga}},{{Mn}})({{As}},{{P}})},}\ }\href {https://doi.org/10.1103/PhysRevB.82.104422} {\bibfield  {journal} {\bibinfo  {journal} {Physical Review B}\ }\textbf {\bibinfo {volume} {82}},\ \bibinfo {pages} {104422} (\bibinfo {year} {2010})}\BibitemShut {NoStop}%
\bibitem [{\citenamefont {Kuszewski}\ \emph {et~al.}(2018)\citenamefont {Kuszewski}, \citenamefont {Duquesne}, \citenamefont {Becerra}, \citenamefont {Lema{\^i}tre}, \citenamefont {Vincent}, \citenamefont {Majrab}, \citenamefont {Margaillan}, \citenamefont {Gourdon},\ and\ \citenamefont {Thevenard}}]{kuszewskiOpticalProbingRayleigh2018}%
  \BibitemOpen
  \bibfield  {author} {\bibinfo {author} {\bibfnamefont {P.}~\bibnamefont {Kuszewski}}, \bibinfo {author} {\bibfnamefont {J.-Y.}\ \bibnamefont {Duquesne}}, \bibinfo {author} {\bibfnamefont {L.}~\bibnamefont {Becerra}}, \bibinfo {author} {\bibfnamefont {A.}~\bibnamefont {Lema{\^i}tre}}, \bibinfo {author} {\bibfnamefont {S.}~\bibnamefont {Vincent}}, \bibinfo {author} {\bibfnamefont {S.}~\bibnamefont {Majrab}}, \bibinfo {author} {\bibfnamefont {F.}~\bibnamefont {Margaillan}}, \bibinfo {author} {\bibfnamefont {C.}~\bibnamefont {Gourdon}},\ and\ \bibinfo {author} {\bibfnamefont {L.}~\bibnamefont {Thevenard}},\ }\bibfield  {title} {\enquote {\bibinfo {title} {Optical probing of rayleigh wave driven magnetoacoustic resonance},}\ }\href {https://doi.org/10.1103/PhysRevApplied.10.034036} {\bibfield  {journal} {\bibinfo  {journal} {Physical Review Applied}\ }\textbf {\bibinfo {volume} {10}},\ \bibinfo {pages} {034036} (\bibinfo {year} {2018})}\BibitemShut {NoStop}%
\bibitem [{\citenamefont {Thevenard}\ \emph {et~al.}(2016)\citenamefont {Thevenard}, \citenamefont {Camara}, \citenamefont {Majrab}, \citenamefont {Bernard}, \citenamefont {Rovillain}, \citenamefont {Lema{\^i}tre}, \citenamefont {Gourdon},\ and\ \citenamefont {Duquesne}}]{thevenardPrecessionalMagnetizationSwitching2016}%
  \BibitemOpen
  \bibfield  {author} {\bibinfo {author} {\bibfnamefont {L.}~\bibnamefont {Thevenard}}, \bibinfo {author} {\bibfnamefont {I.~S.}\ \bibnamefont {Camara}}, \bibinfo {author} {\bibfnamefont {S.}~\bibnamefont {Majrab}}, \bibinfo {author} {\bibfnamefont {M.}~\bibnamefont {Bernard}}, \bibinfo {author} {\bibfnamefont {P.}~\bibnamefont {Rovillain}}, \bibinfo {author} {\bibfnamefont {A.}~\bibnamefont {Lema{\^i}tre}}, \bibinfo {author} {\bibfnamefont {C.}~\bibnamefont {Gourdon}},\ and\ \bibinfo {author} {\bibfnamefont {J.-Y.}\ \bibnamefont {Duquesne}},\ }\bibfield  {title} {\enquote {\bibinfo {title} {Precessional magnetization switching by a surface acoustic wave},}\ }\href {https://doi.org/10.1103/PhysRevB.93.134430} {\bibfield  {journal} {\bibinfo  {journal} {Physical Review B}\ }\textbf {\bibinfo {volume} {93}},\ \bibinfo {pages} {134430} (\bibinfo {year} {2016})}\BibitemShut {NoStop}%
\bibitem [{\citenamefont {Hashimoto}\ and\ \citenamefont {Ochiai}(1990)}]{hashimotoCoPtCo1990}%
  \BibitemOpen
  \bibfield  {author} {\bibinfo {author} {\bibfnamefont {S.}~\bibnamefont {Hashimoto}}\ and\ \bibinfo {author} {\bibfnamefont {Y.}~\bibnamefont {Ochiai}},\ }\bibfield  {title} {\enquote {\bibinfo {title} {{{Co}}/{{Pt}} and {{Co}}/{{Pd}} multilayers as magneto-optical recording materials},}\ }\href {https://doi.org/10.1016/S0304-8853(97)90031-7} {\bibfield  {journal} {\bibinfo  {journal} {Journal of Magnetism and Magnetic Materials}\ }\textbf {\bibinfo {volume} {88}},\ \bibinfo {pages} {211--226} (\bibinfo {year} {1990})}\BibitemShut {NoStop}%
\bibitem [{\citenamefont {Hashimoto}, \citenamefont {Ochiai},\ and\ \citenamefont {Aso}(1990)}]{hashimotoFilmThicknessDependence1990}%
  \BibitemOpen
  \bibfield  {author} {\bibinfo {author} {\bibfnamefont {S.}~\bibnamefont {Hashimoto}}, \bibinfo {author} {\bibfnamefont {Y.}~\bibnamefont {Ochiai}},\ and\ \bibinfo {author} {\bibfnamefont {K.}~\bibnamefont {Aso}},\ }\bibfield  {title} {\enquote {\bibinfo {title} {Film thickness dependence of magneto-optical and magnetic properties in {{Co/Pt}} and {{Co}}/{{Pd}} multilayers},}\ }\href {https://doi.org/10.1063/1.344921} {\bibfield  {journal} {\bibinfo  {journal} {Journal of Applied Physics}\ }\textbf {\bibinfo {volume} {67}},\ \bibinfo {pages} {4429--4431} (\bibinfo {year} {1990})}\BibitemShut {NoStop}%
\bibitem [{\citenamefont {Shim}\ \emph {et~al.}(2017)\citenamefont {Shim}, \citenamefont {Ali~Syed}, \citenamefont {Kim}, \citenamefont {Lee}, \citenamefont {Park}, \citenamefont {Jeong}, \citenamefont {Kim},\ and\ \citenamefont {Eon~Kim}}]{shimUltrafastGiantMagnetic2017}%
  \BibitemOpen
  \bibfield  {author} {\bibinfo {author} {\bibfnamefont {J.-H.}\ \bibnamefont {Shim}}, \bibinfo {author} {\bibfnamefont {A.}~\bibnamefont {Ali~Syed}}, \bibinfo {author} {\bibfnamefont {C.-H.}\ \bibnamefont {Kim}}, \bibinfo {author} {\bibfnamefont {K.~M.}\ \bibnamefont {Lee}}, \bibinfo {author} {\bibfnamefont {S.-Y.}\ \bibnamefont {Park}}, \bibinfo {author} {\bibfnamefont {J.-R.}\ \bibnamefont {Jeong}}, \bibinfo {author} {\bibfnamefont {D.-H.}\ \bibnamefont {Kim}},\ and\ \bibinfo {author} {\bibfnamefont {D.}~\bibnamefont {Eon~Kim}},\ }\bibfield  {title} {\enquote {\bibinfo {title} {Ultrafast giant magnetic cooling effect in ferromagnetic {{Co}}/{{Pt}} multilayers},}\ }\href {https://doi.org/10.1038/s41467-017-00816-w} {\bibfield  {journal} {\bibinfo  {journal} {Nature Communications}\ }\textbf {\bibinfo {volume} {8}},\ \bibinfo {pages} {796} (\bibinfo {year} {2017})}\BibitemShut {NoStop}%
\bibitem [{\citenamefont {Kim}\ \emph {et~al.}(2016)\citenamefont {Kim}, \citenamefont {Shim}, \citenamefont {Lee}, \citenamefont {Jeong}, \citenamefont {Kim},\ and\ \citenamefont {Kim}}]{kimCoherentPhononControl2016}%
  \BibitemOpen
  \bibfield  {author} {\bibinfo {author} {\bibfnamefont {C.~H.}\ \bibnamefont {Kim}}, \bibinfo {author} {\bibfnamefont {J.-H.}\ \bibnamefont {Shim}}, \bibinfo {author} {\bibfnamefont {K.~M.}\ \bibnamefont {Lee}}, \bibinfo {author} {\bibfnamefont {J.-R.}\ \bibnamefont {Jeong}}, \bibinfo {author} {\bibfnamefont {D.-H.}\ \bibnamefont {Kim}},\ and\ \bibinfo {author} {\bibfnamefont {D.~E.}\ \bibnamefont {Kim}},\ }\bibfield  {title} {\enquote {\bibinfo {title} {Coherent phonon control via electron-lattice interaction in ferromagnetic {{Co}}/{{Pt}} multilayers},}\ }\href {https://doi.org/10.1038/srep22054} {\bibfield  {journal} {\bibinfo  {journal} {Scientific Reports}\ }\textbf {\bibinfo {volume} {6}},\ \bibinfo {pages} {1--8} (\bibinfo {year} {2016})}\BibitemShut {NoStop}%
\bibitem [{\citenamefont {Gu}\ \emph {et~al.}(2024)\citenamefont {Gu}, \citenamefont {Xu}, \citenamefont {Delodovici}, \citenamefont {Carcan}, \citenamefont {Khiari}, \citenamefont {Vaudel}, \citenamefont {Juv{\'e}}, \citenamefont {Weber}, \citenamefont {Poirier}, \citenamefont {Nandi}, \citenamefont {Xu}, \citenamefont {Gusev}, \citenamefont {Bellaiche}, \citenamefont {Laulh{\'e}}, \citenamefont {Jaouen}, \citenamefont {Manuel}, \citenamefont {Dkhil}, \citenamefont {Paillard}, \citenamefont {Yedra}, \citenamefont {Bouyanfif},\ and\ \citenamefont {Ruello}}]{guSuperordersTerahertzAcoustic2024}%
  \BibitemOpen
  \bibfield  {author} {\bibinfo {author} {\bibfnamefont {R.}~\bibnamefont {Gu}}, \bibinfo {author} {\bibfnamefont {R.}~\bibnamefont {Xu}}, \bibinfo {author} {\bibfnamefont {F.}~\bibnamefont {Delodovici}}, \bibinfo {author} {\bibfnamefont {B.}~\bibnamefont {Carcan}}, \bibinfo {author} {\bibfnamefont {M.}~\bibnamefont {Khiari}}, \bibinfo {author} {\bibfnamefont {G.}~\bibnamefont {Vaudel}}, \bibinfo {author} {\bibfnamefont {V.}~\bibnamefont {Juv{\'e}}}, \bibinfo {author} {\bibfnamefont {M.~C.}\ \bibnamefont {Weber}}, \bibinfo {author} {\bibfnamefont {A.}~\bibnamefont {Poirier}}, \bibinfo {author} {\bibfnamefont {P.}~\bibnamefont {Nandi}}, \bibinfo {author} {\bibfnamefont {B.}~\bibnamefont {Xu}}, \bibinfo {author} {\bibfnamefont {V.~E.}\ \bibnamefont {Gusev}}, \bibinfo {author} {\bibfnamefont {L.}~\bibnamefont {Bellaiche}}, \bibinfo {author} {\bibfnamefont {C.}~\bibnamefont {Laulh{\'e}}}, \bibinfo {author} {\bibfnamefont {N.}~\bibnamefont {Jaouen}}, \bibinfo {author} {\bibfnamefont {P.}~\bibnamefont {Manuel}},
  \bibinfo {author} {\bibfnamefont {B.}~\bibnamefont {Dkhil}}, \bibinfo {author} {\bibfnamefont {C.}~\bibnamefont {Paillard}}, \bibinfo {author} {\bibfnamefont {L.}~\bibnamefont {Yedra}}, \bibinfo {author} {\bibfnamefont {H.}~\bibnamefont {Bouyanfif}},\ and\ \bibinfo {author} {\bibfnamefont {P.}~\bibnamefont {Ruello}},\ }\bibfield  {title} {\enquote {\bibinfo {title} {Superorders and terahertz acoustic modes in multiferroic {{BiFeO}}$_3$/{{LaFeO}}$_3$ superlattices},}\ }\href {https://doi.org/10.1063/5.0203076} {\bibfield  {journal} {\bibinfo  {journal} {Applied Physics Reviews}\ }\textbf {\bibinfo {volume} {11}},\ \bibinfo {pages} {041415} (\bibinfo {year} {2024})}\BibitemShut {NoStop}%
\bibitem [{\citenamefont {Huynh}, \citenamefont {Perrin},\ and\ \citenamefont {Lema{\^i}tre}(2015)}]{huynhSemiconductorSuperlatticesTool2015}%
  \BibitemOpen
  \bibfield  {author} {\bibinfo {author} {\bibfnamefont {A.}~\bibnamefont {Huynh}}, \bibinfo {author} {\bibfnamefont {B.}~\bibnamefont {Perrin}},\ and\ \bibinfo {author} {\bibfnamefont {A.}~\bibnamefont {Lema{\^i}tre}},\ }\bibfield  {title} {\enquote {\bibinfo {title} {Semiconductor superlattices: {{A}} tool for terahertz acoustics},}\ }\href {https://doi.org/10.1016/j.ultras.2014.07.009} {\bibfield  {journal} {\bibinfo  {journal} {Ultrasonics}\ }\textbf {\bibinfo {volume} {56}},\ \bibinfo {pages} {66--79} (\bibinfo {year} {2015})}\BibitemShut {NoStop}%
\bibitem [{\citenamefont {{Lanzillotti-Kimura}}\ \emph {et~al.}(2006)\citenamefont {{Lanzillotti-Kimura}}, \citenamefont {Fainstein}, \citenamefont {Lema{\^i}tre},\ and\ \citenamefont {Jusserand}}]{lanzillotti-kimuraNanowaveDevicesTerahertz2006}%
  \BibitemOpen
  \bibfield  {author} {\bibinfo {author} {\bibfnamefont {N.~D.}\ \bibnamefont {{Lanzillotti-Kimura}}}, \bibinfo {author} {\bibfnamefont {A.}~\bibnamefont {Fainstein}}, \bibinfo {author} {\bibfnamefont {A.}~\bibnamefont {Lema{\^i}tre}},\ and\ \bibinfo {author} {\bibfnamefont {B.}~\bibnamefont {Jusserand}},\ }\bibfield  {title} {\enquote {\bibinfo {title} {Nanowave devices for terahertz acoustic phonons},}\ }\href {https://doi.org/10.1063/1.2178415} {\bibfield  {journal} {\bibinfo  {journal} {Applied Physics Letters}\ }\textbf {\bibinfo {volume} {88}},\ \bibinfo {pages} {083113} (\bibinfo {year} {2006})}\BibitemShut {NoStop}%
\bibitem [{\citenamefont {{Lanzillotti-Kimura}}\ \emph {et~al.}(2007)\citenamefont {{Lanzillotti-Kimura}}, \citenamefont {Fainstein}, \citenamefont {Jusserand}, \citenamefont {Lema{\^i}tre}, \citenamefont {Mauguin},\ and\ \citenamefont {Largeau}}]{lanzillotti-kimuraAcousticPhononNanowave2007}%
  \BibitemOpen
  \bibfield  {author} {\bibinfo {author} {\bibfnamefont {N.~D.}\ \bibnamefont {{Lanzillotti-Kimura}}}, \bibinfo {author} {\bibfnamefont {A.}~\bibnamefont {Fainstein}}, \bibinfo {author} {\bibfnamefont {B.}~\bibnamefont {Jusserand}}, \bibinfo {author} {\bibfnamefont {A.}~\bibnamefont {Lema{\^i}tre}}, \bibinfo {author} {\bibfnamefont {O.}~\bibnamefont {Mauguin}},\ and\ \bibinfo {author} {\bibfnamefont {L.}~\bibnamefont {Largeau}},\ }\bibfield  {title} {\enquote {\bibinfo {title} {Acoustic phonon nanowave devices based on aperiodic multilayers: {{Experiments}} and theory},}\ }\href {https://doi.org/10.1103/PhysRevB.76.174301} {\bibfield  {journal} {\bibinfo  {journal} {Physical Review B}\ }\textbf {\bibinfo {volume} {76}},\ \bibinfo {pages} {174301} (\bibinfo {year} {2007})}\BibitemShut {NoStop}%
\bibitem [{\citenamefont {Narayanamurti}\ \emph {et~al.}(1979)\citenamefont {Narayanamurti}, \citenamefont {St{\"o}rmer}, \citenamefont {Chin}, \citenamefont {Gossard},\ and\ \citenamefont {Wiegmann}}]{narayanamurtiSelectiveTransmissionHighFrequency1979}%
  \BibitemOpen
  \bibfield  {author} {\bibinfo {author} {\bibfnamefont {V.}~\bibnamefont {Narayanamurti}}, \bibinfo {author} {\bibfnamefont {H.~L.}\ \bibnamefont {St{\"o}rmer}}, \bibinfo {author} {\bibfnamefont {M.~A.}\ \bibnamefont {Chin}}, \bibinfo {author} {\bibfnamefont {A.~C.}\ \bibnamefont {Gossard}},\ and\ \bibinfo {author} {\bibfnamefont {W.}~\bibnamefont {Wiegmann}},\ }\bibfield  {title} {\enquote {\bibinfo {title} {Selective {{Transmission}} of {{High-Frequency Phonons}} by a {{Superlattice}}: {{The}} "{{Dielectric}}" {{Phonon Filter}}},}\ }\href {https://doi.org/10.1103/PhysRevLett.43.2012} {\bibfield  {journal} {\bibinfo  {journal} {Physical Review Letters}\ }\textbf {\bibinfo {volume} {43}},\ \bibinfo {pages} {2012--2016} (\bibinfo {year} {1979})}\BibitemShut {NoStop}%
\bibitem [{\citenamefont {Ortiz}\ \emph {et~al.}(2021)\citenamefont {Ortiz}, \citenamefont {Priya}, \citenamefont {Rodriguez}, \citenamefont {Lemaitre}, \citenamefont {Esmann},\ and\ \citenamefont {{Lanzillotti-Kimura}}}]{ortizTopologicalOpticalPhononic2021}%
  \BibitemOpen
  \bibfield  {author} {\bibinfo {author} {\bibfnamefont {O.}~\bibnamefont {Ortiz}}, \bibinfo {author} {\bibfnamefont {P.}~\bibnamefont {Priya}}, \bibinfo {author} {\bibfnamefont {A.}~\bibnamefont {Rodriguez}}, \bibinfo {author} {\bibfnamefont {A.}~\bibnamefont {Lemaitre}}, \bibinfo {author} {\bibfnamefont {M.}~\bibnamefont {Esmann}},\ and\ \bibinfo {author} {\bibfnamefont {N.~D.}\ \bibnamefont {{Lanzillotti-Kimura}}},\ }\bibfield  {title} {\enquote {\bibinfo {title} {Topological optical and phononic interface mode by simultaneous band inversion},}\ }\href {https://doi.org/10.1364/OPTICA.411945} {\bibfield  {journal} {\bibinfo  {journal} {Optica}\ }\textbf {\bibinfo {volume} {8}},\ \bibinfo {pages} {598} (\bibinfo {year} {2021})}\BibitemShut {NoStop}%
\bibitem [{\citenamefont {Ort{\'i}z}, \citenamefont {Esmann},\ and\ \citenamefont {{Lanzillotti-Kimura}}(2019)}]{ortizPhononEngineeringSuperlattices2019}%
  \BibitemOpen
  \bibfield  {author} {\bibinfo {author} {\bibfnamefont {O.}~\bibnamefont {Ort{\'i}z}}, \bibinfo {author} {\bibfnamefont {M.}~\bibnamefont {Esmann}},\ and\ \bibinfo {author} {\bibfnamefont {N.~D.}\ \bibnamefont {{Lanzillotti-Kimura}}},\ }\bibfield  {title} {\enquote {\bibinfo {title} {Phonon engineering with superlattices: {{Generalized}} nanomechanical potentials},}\ }\href {https://doi.org/10.1103/PhysRevB.100.085430} {\bibfield  {journal} {\bibinfo  {journal} {Physical Review B}\ }\textbf {\bibinfo {volume} {100}},\ \bibinfo {pages} {085430} (\bibinfo {year} {2019})}\BibitemShut {NoStop}%
\bibitem [{\citenamefont {Arregui}\ \emph {et~al.}(2019{\natexlab{a}})\citenamefont {Arregui}, \citenamefont {Ort{\'i}z}, \citenamefont {Esmann}, \citenamefont {{Sotomayor-Torres}}, \citenamefont {{Gomez-Carbonell}}, \citenamefont {Mauguin}, \citenamefont {Perrin}, \citenamefont {Lema{\^i}tre}, \citenamefont {Garc{\'i}a},\ and\ \citenamefont {{Lanzillotti-Kimura}}}]{arreguiCoherentGenerationDetection2019}%
  \BibitemOpen
  \bibfield  {author} {\bibinfo {author} {\bibfnamefont {G.}~\bibnamefont {Arregui}}, \bibinfo {author} {\bibfnamefont {O.}~\bibnamefont {Ort{\'i}z}}, \bibinfo {author} {\bibfnamefont {M.}~\bibnamefont {Esmann}}, \bibinfo {author} {\bibfnamefont {C.~M.}\ \bibnamefont {{Sotomayor-Torres}}}, \bibinfo {author} {\bibfnamefont {C.}~\bibnamefont {{Gomez-Carbonell}}}, \bibinfo {author} {\bibfnamefont {O.}~\bibnamefont {Mauguin}}, \bibinfo {author} {\bibfnamefont {B.}~\bibnamefont {Perrin}}, \bibinfo {author} {\bibfnamefont {A.}~\bibnamefont {Lema{\^i}tre}}, \bibinfo {author} {\bibfnamefont {P.~D.}\ \bibnamefont {Garc{\'i}a}},\ and\ \bibinfo {author} {\bibfnamefont {N.~D.}\ \bibnamefont {{Lanzillotti-Kimura}}},\ }\bibfield  {title} {\enquote {\bibinfo {title} {Coherent generation and detection of acoustic phonons in topological nanocavities},}\ }\href {https://doi.org/10.1063/1.5082728} {\bibfield  {journal} {\bibinfo  {journal} {APL Photonics}\ }\textbf {\bibinfo {volume} {4}},\ \bibinfo {pages} {030805} (\bibinfo
  {year} {2019}{\natexlab{a}})}\BibitemShut {NoStop}%
\bibitem [{\citenamefont {Esmann}\ \emph {et~al.}(2018)\citenamefont {Esmann}, \citenamefont {Lamberti}, \citenamefont {Senellart}, \citenamefont {Favero}, \citenamefont {Krebs}, \citenamefont {Lanco}, \citenamefont {Gomez~Carbonell}, \citenamefont {Lema{\^i}tre},\ and\ \citenamefont {{Lanzillotti-Kimura}}}]{esmannTopologicalNanophononicStates2018}%
  \BibitemOpen
  \bibfield  {author} {\bibinfo {author} {\bibfnamefont {M.}~\bibnamefont {Esmann}}, \bibinfo {author} {\bibfnamefont {F.~R.}\ \bibnamefont {Lamberti}}, \bibinfo {author} {\bibfnamefont {P.}~\bibnamefont {Senellart}}, \bibinfo {author} {\bibfnamefont {I.}~\bibnamefont {Favero}}, \bibinfo {author} {\bibfnamefont {O.}~\bibnamefont {Krebs}}, \bibinfo {author} {\bibfnamefont {L.}~\bibnamefont {Lanco}}, \bibinfo {author} {\bibfnamefont {C.}~\bibnamefont {Gomez~Carbonell}}, \bibinfo {author} {\bibfnamefont {A.}~\bibnamefont {Lema{\^i}tre}},\ and\ \bibinfo {author} {\bibfnamefont {N.~D.}\ \bibnamefont {{Lanzillotti-Kimura}}},\ }\bibfield  {title} {\enquote {\bibinfo {title} {Topological nanophononic states by band inversion},}\ }\href {https://doi.org/10.1103/PhysRevB.97.155422} {\bibfield  {journal} {\bibinfo  {journal} {Physical Review B}\ }\textbf {\bibinfo {volume} {97}},\ \bibinfo {pages} {155422} (\bibinfo {year} {2018})}\BibitemShut {NoStop}%
\bibitem [{\citenamefont {Rodriguez}\ \emph {et~al.}(2023)\citenamefont {Rodriguez}, \citenamefont {Papatryfonos}, \citenamefont {Cardozo De~Oliveira},\ and\ \citenamefont {{Lanzillotti-Kimura}}}]{rodriguezTopologicalNanophononicInterface2023}%
  \BibitemOpen
  \bibfield  {author} {\bibinfo {author} {\bibfnamefont {A.}~\bibnamefont {Rodriguez}}, \bibinfo {author} {\bibfnamefont {K.}~\bibnamefont {Papatryfonos}}, \bibinfo {author} {\bibfnamefont {E.~R.}\ \bibnamefont {Cardozo De~Oliveira}},\ and\ \bibinfo {author} {\bibfnamefont {N.~D.}\ \bibnamefont {{Lanzillotti-Kimura}}},\ }\bibfield  {title} {\enquote {\bibinfo {title} {Topological nanophononic interface states using high-order bandgaps in the one-dimensional {{Su-Schrieffer-Heeger}} model},}\ }\href {https://doi.org/10.1103/PhysRevB.108.205301} {\bibfield  {journal} {\bibinfo  {journal} {Physical Review B}\ }\textbf {\bibinfo {volume} {108}},\ \bibinfo {pages} {205301} (\bibinfo {year} {2023})}\BibitemShut {NoStop}%
\bibitem [{\citenamefont {{Lanzillotti-Kimura}}\ \emph {et~al.}(2010{\natexlab{a}})\citenamefont {{Lanzillotti-Kimura}}, \citenamefont {Fainstein}, \citenamefont {Perrin}, \citenamefont {Jusserand}, \citenamefont {Mauguin}, \citenamefont {Largeau},\ and\ \citenamefont {Lema{\^i}tre}}]{lanzillotti-kimuraBlochOscillationsTHz2010}%
  \BibitemOpen
  \bibfield  {author} {\bibinfo {author} {\bibfnamefont {N.~D.}\ \bibnamefont {{Lanzillotti-Kimura}}}, \bibinfo {author} {\bibfnamefont {A.}~\bibnamefont {Fainstein}}, \bibinfo {author} {\bibfnamefont {B.}~\bibnamefont {Perrin}}, \bibinfo {author} {\bibfnamefont {B.}~\bibnamefont {Jusserand}}, \bibinfo {author} {\bibfnamefont {O.}~\bibnamefont {Mauguin}}, \bibinfo {author} {\bibfnamefont {L.}~\bibnamefont {Largeau}},\ and\ \bibinfo {author} {\bibfnamefont {A.}~\bibnamefont {Lema{\^i}tre}},\ }\bibfield  {title} {\enquote {\bibinfo {title} {Bloch {{Oscillations}} of {{THz Acoustic Phonons}} in {{Coupled Nanocavity Structures}}},}\ }\href {https://doi.org/10.1103/PhysRevLett.104.197402} {\bibfield  {journal} {\bibinfo  {journal} {Physical Review Letters}\ }\textbf {\bibinfo {volume} {104}},\ \bibinfo {pages} {197402} (\bibinfo {year} {2010}{\natexlab{a}})}\BibitemShut {NoStop}%
\bibitem [{\citenamefont {Arregui}\ \emph {et~al.}(2019{\natexlab{b}})\citenamefont {Arregui}, \citenamefont {{Lanzillotti-Kimura}}, \citenamefont {{Sotomayor-Torres}},\ and\ \citenamefont {Garc{\'i}a}}]{arreguiAndersonPhotonPhononColocalization2019}%
  \BibitemOpen
  \bibfield  {author} {\bibinfo {author} {\bibfnamefont {G.}~\bibnamefont {Arregui}}, \bibinfo {author} {\bibfnamefont {N.~D.}\ \bibnamefont {{Lanzillotti-Kimura}}}, \bibinfo {author} {\bibfnamefont {C.~M.}\ \bibnamefont {{Sotomayor-Torres}}},\ and\ \bibinfo {author} {\bibfnamefont {P.~D.}\ \bibnamefont {Garc{\'i}a}},\ }\bibfield  {title} {\enquote {\bibinfo {title} {Anderson {{Photon-Phonon Colocalization}} in {{Certain Random Superlattices}}},}\ }\href {https://doi.org/10.1103/PhysRevLett.122.043903} {\bibfield  {journal} {\bibinfo  {journal} {Physical Review Letters}\ }\textbf {\bibinfo {volume} {122}},\ \bibinfo {pages} {043903} (\bibinfo {year} {2019}{\natexlab{b}})}\BibitemShut {NoStop}%
\bibitem [{\citenamefont {{Lanzillotti-Kimura}}\ \emph {et~al.}(2010{\natexlab{b}})\citenamefont {{Lanzillotti-Kimura}}, \citenamefont {Fainstein}, \citenamefont {Perrin}, \citenamefont {Jusserand}, \citenamefont {Soukiassian}, \citenamefont {Xi},\ and\ \citenamefont {Schlom}}]{lanzillotti-kimuraEnhancementInhibitionCoherent2010}%
  \BibitemOpen
  \bibfield  {author} {\bibinfo {author} {\bibfnamefont {N.~D.}\ \bibnamefont {{Lanzillotti-Kimura}}}, \bibinfo {author} {\bibfnamefont {A.}~\bibnamefont {Fainstein}}, \bibinfo {author} {\bibfnamefont {B.}~\bibnamefont {Perrin}}, \bibinfo {author} {\bibfnamefont {B.}~\bibnamefont {Jusserand}}, \bibinfo {author} {\bibfnamefont {A.}~\bibnamefont {Soukiassian}}, \bibinfo {author} {\bibfnamefont {X.~X.}\ \bibnamefont {Xi}},\ and\ \bibinfo {author} {\bibfnamefont {D.~G.}\ \bibnamefont {Schlom}},\ }\bibfield  {title} {\enquote {\bibinfo {title} {Enhancement and {{Inhibition}} of {{Coherent Phonon Emission}} of a {{Ni Film}} in a {{BaTiO}}$_3$ / {{SrTiO}}$_3$ {{Cavity}}},}\ }\href {https://doi.org/10.1103/PhysRevLett.104.187402} {\bibfield  {journal} {\bibinfo  {journal} {Physical Review Letters}\ }\textbf {\bibinfo {volume} {104}},\ \bibinfo {pages} {187402} (\bibinfo {year} {2010}{\natexlab{b}})}\BibitemShut {NoStop}%
\bibitem [{\citenamefont {Gomopoulos}\ \emph {et~al.}(2010)\citenamefont {Gomopoulos}, \citenamefont {Maschke}, \citenamefont {Koh}, \citenamefont {Thomas}, \citenamefont {Tremel}, \citenamefont {Butt},\ and\ \citenamefont {Fytas}}]{gomopoulosOneDimensionalHypersonicPhononic2010a}%
  \BibitemOpen
  \bibfield  {author} {\bibinfo {author} {\bibfnamefont {N.}~\bibnamefont {Gomopoulos}}, \bibinfo {author} {\bibfnamefont {D.}~\bibnamefont {Maschke}}, \bibinfo {author} {\bibfnamefont {C.~Y.}\ \bibnamefont {Koh}}, \bibinfo {author} {\bibfnamefont {E.~L.}\ \bibnamefont {Thomas}}, \bibinfo {author} {\bibfnamefont {W.}~\bibnamefont {Tremel}}, \bibinfo {author} {\bibfnamefont {H.-J.}\ \bibnamefont {Butt}},\ and\ \bibinfo {author} {\bibfnamefont {G.}~\bibnamefont {Fytas}},\ }\bibfield  {title} {\enquote {\bibinfo {title} {One-dimensional hypersonic phononic crystals},}\ }\href {https://doi.org/10.1021/nl903959r} {\bibfield  {journal} {\bibinfo  {journal} {Nano Letters}\ }\textbf {\bibinfo {volume} {10}},\ \bibinfo {pages} {980--984} (\bibinfo {year} {2010})}\BibitemShut {NoStop}%
\bibitem [{\citenamefont {Schneider}\ \emph {et~al.}(2013)\citenamefont {Schneider}, \citenamefont {Liaqat}, \citenamefont {El~Boudouti}, \citenamefont {El~Abouti}, \citenamefont {Tremel}, \citenamefont {Butt}, \citenamefont {{Djafari-Rouhani}},\ and\ \citenamefont {Fytas}}]{schneiderDefectControlledHypersoundPropagation2013}%
  \BibitemOpen
  \bibfield  {author} {\bibinfo {author} {\bibfnamefont {D.}~\bibnamefont {Schneider}}, \bibinfo {author} {\bibfnamefont {F.}~\bibnamefont {Liaqat}}, \bibinfo {author} {\bibfnamefont {E.~H.}\ \bibnamefont {El~Boudouti}}, \bibinfo {author} {\bibfnamefont {O.}~\bibnamefont {El~Abouti}}, \bibinfo {author} {\bibfnamefont {W.}~\bibnamefont {Tremel}}, \bibinfo {author} {\bibfnamefont {H.-J.}\ \bibnamefont {Butt}}, \bibinfo {author} {\bibfnamefont {B.}~\bibnamefont {{Djafari-Rouhani}}},\ and\ \bibinfo {author} {\bibfnamefont {G.}~\bibnamefont {Fytas}},\ }\bibfield  {title} {\enquote {\bibinfo {title} {Defect-controlled hypersound propagation in hybrid superlattices},}\ }\href {https://doi.org/10.1103/PhysRevLett.111.164301} {\bibfield  {journal} {\bibinfo  {journal} {Physical Review Letters}\ }\textbf {\bibinfo {volume} {111}},\ \bibinfo {pages} {164301} (\bibinfo {year} {2013})}\BibitemShut {NoStop}%
\bibitem [{\citenamefont {Ezzahri}\ \emph {et~al.}(2007)\citenamefont {Ezzahri}, \citenamefont {Grauby}, \citenamefont {Rampnoux}, \citenamefont {Michel}, \citenamefont {Pernot}, \citenamefont {Claeys}, \citenamefont {Dilhaire}, \citenamefont {Rossignol}, \citenamefont {Zeng},\ and\ \citenamefont {Shakouri}}]{ezzahriCoherentPhononsSi2007}%
  \BibitemOpen
  \bibfield  {author} {\bibinfo {author} {\bibfnamefont {Y.}~\bibnamefont {Ezzahri}}, \bibinfo {author} {\bibfnamefont {S.}~\bibnamefont {Grauby}}, \bibinfo {author} {\bibfnamefont {J.~M.}\ \bibnamefont {Rampnoux}}, \bibinfo {author} {\bibfnamefont {H.}~\bibnamefont {Michel}}, \bibinfo {author} {\bibfnamefont {G.}~\bibnamefont {Pernot}}, \bibinfo {author} {\bibfnamefont {W.}~\bibnamefont {Claeys}}, \bibinfo {author} {\bibfnamefont {S.}~\bibnamefont {Dilhaire}}, \bibinfo {author} {\bibfnamefont {C.}~\bibnamefont {Rossignol}}, \bibinfo {author} {\bibfnamefont {G.}~\bibnamefont {Zeng}},\ and\ \bibinfo {author} {\bibfnamefont {A.}~\bibnamefont {Shakouri}},\ }\bibfield  {title} {\enquote {\bibinfo {title} {Coherent phonons in {{Si}}/{{SiGe}} superlattices},}\ }\href {https://doi.org/10.1103/PhysRevB.75.195309} {\bibfield  {journal} {\bibinfo  {journal} {Physical Review B}\ }\textbf {\bibinfo {volume} {75}},\ \bibinfo {pages} {195309} (\bibinfo {year} {2007})}\BibitemShut {NoStop}%
\bibitem [{\citenamefont {Wilson}(2018)}]{wilsonEvidenceTerahertzAcoustic2018}%
  \BibitemOpen
  \bibfield  {author} {\bibinfo {author} {\bibfnamefont {T.~E.}\ \bibnamefont {Wilson}},\ }\bibfield  {title} {\enquote {\bibinfo {title} {Evidence for terahertz acoustic phonon parametric oscillator based on acousto-optic degenerate four-wave mixing in a silicon doping superlattice},}\ }\href {https://doi.org/10.1103/PhysRevB.98.220304} {\bibfield  {journal} {\bibinfo  {journal} {Physical Review B}\ }\textbf {\bibinfo {volume} {98}},\ \bibinfo {pages} {220304} (\bibinfo {year} {2018})}\BibitemShut {NoStop}%
\bibitem [{\citenamefont {Parsons}\ and\ \citenamefont {Andrews}(2012)}]{parsonsBrillouinScatteringPorous2012}%
  \BibitemOpen
  \bibfield  {author} {\bibinfo {author} {\bibfnamefont {L.~C.}\ \bibnamefont {Parsons}}\ and\ \bibinfo {author} {\bibfnamefont {G.~T.}\ \bibnamefont {Andrews}},\ }\bibfield  {title} {\enquote {\bibinfo {title} {Brillouin scattering from porous silicon-based optical bragg mirrors},}\ }\href {https://doi.org/10.1063/1.4730617} {\bibfield  {journal} {\bibinfo  {journal} {Journal of Applied Physics}\ }\textbf {\bibinfo {volume} {111}},\ \bibinfo {pages} {123521} (\bibinfo {year} {2012})}\BibitemShut {NoStop}%
\bibitem [{\citenamefont {Perrin}\ \emph {et~al.}(1996)\citenamefont {Perrin}, \citenamefont {Bonello}, \citenamefont {Jeannet},\ and\ \citenamefont {Romatet}}]{perrinPicosecondUltrasonicsStudy1996}%
  \BibitemOpen
  \bibfield  {author} {\bibinfo {author} {\bibfnamefont {B.}~\bibnamefont {Perrin}}, \bibinfo {author} {\bibfnamefont {B.}~\bibnamefont {Bonello}}, \bibinfo {author} {\bibfnamefont {J.-C.}\ \bibnamefont {Jeannet}},\ and\ \bibinfo {author} {\bibfnamefont {E.}~\bibnamefont {Romatet}},\ }\bibfield  {title} {\enquote {\bibinfo {title} {Picosecond ultrasonics study of metallic multilayers},}\ }\href {https://doi.org/10.1016/0921-4526(95)00852-7} {\bibfield  {journal} {\bibinfo  {journal} {Physica B: Condensed Matter}\ }\textbf {\bibinfo {volume} {219--220}},\ \bibinfo {pages} {681--683} (\bibinfo {year} {1996})}\BibitemShut {NoStop}%
\bibitem [{\citenamefont {Cardozo De~Oliveira}\ \emph {et~al.}(2023)\citenamefont {Cardozo De~Oliveira}, \citenamefont {Vensaus}, \citenamefont {{Soler-Illia}},\ and\ \citenamefont {{Lanzillotti-Kimura}}}]{cardozodeoliveiraDesignCosteffectiveEnvironmentresponsive2023}%
  \BibitemOpen
  \bibfield  {author} {\bibinfo {author} {\bibfnamefont {E.~R.}\ \bibnamefont {Cardozo De~Oliveira}}, \bibinfo {author} {\bibfnamefont {P.}~\bibnamefont {Vensaus}}, \bibinfo {author} {\bibfnamefont {G.~J. A.~A.}\ \bibnamefont {{Soler-Illia}}},\ and\ \bibinfo {author} {\bibfnamefont {N.~D.}\ \bibnamefont {{Lanzillotti-Kimura}}},\ }\bibfield  {title} {\enquote {\bibinfo {title} {Design of cost-effective environment-responsive nanoacoustic devices based on mesoporous thin films},}\ }\href {https://doi.org/10.1364/OME.504926} {\bibfield  {journal} {\bibinfo  {journal} {Optical Materials Express}\ }\textbf {\bibinfo {volume} {13}},\ \bibinfo {pages} {3715} (\bibinfo {year} {2023})}\BibitemShut {NoStop}%
\bibitem [{\citenamefont {{Priya}}, \citenamefont {{Cardozo de Oliveira}},\ and\ \citenamefont {{Lanzillotti-Kimura}}(2023)}]{priyaPerspectivesHighfrequencyNanomechanics2023}%
  \BibitemOpen
  \bibfield  {author} {\bibinfo {author} {\bibnamefont {{Priya}}}, \bibinfo {author} {\bibfnamefont {E.~R.}\ \bibnamefont {{Cardozo de Oliveira}}},\ and\ \bibinfo {author} {\bibfnamefont {N.~D.}\ \bibnamefont {{Lanzillotti-Kimura}}},\ }\bibfield  {title} {\enquote {\bibinfo {title} {Perspectives on high-frequency nanomechanics, nanoacoustics, and nanophononics},}\ }\href {https://doi.org/10.1063/5.0142925} {\bibfield  {journal} {\bibinfo  {journal} {Applied Physics Letters}\ }\textbf {\bibinfo {volume} {122}},\ \bibinfo {pages} {140501} (\bibinfo {year} {2023})}\BibitemShut {NoStop}%
\bibitem [{\citenamefont {Quinteros}\ \emph {et~al.}(2018)\citenamefont {Quinteros}, \citenamefont {Bustingorry}, \citenamefont {Curiale},\ and\ \citenamefont {Granada}}]{QuinterosAPL}%
  \BibitemOpen
  \bibfield  {author} {\bibinfo {author} {\bibfnamefont {C.~P.}\ \bibnamefont {Quinteros}}, \bibinfo {author} {\bibfnamefont {S.}~\bibnamefont {Bustingorry}}, \bibinfo {author} {\bibfnamefont {J.}~\bibnamefont {Curiale}},\ and\ \bibinfo {author} {\bibfnamefont {M.}~\bibnamefont {Granada}},\ }\bibfield  {title} {\enquote {\bibinfo {title} {Correlation between domain wall creep parameters of thin ferromagnetic films},}\ }\href {https://doi.org/10.1063/1.5026702} {\bibfield  {journal} {\bibinfo  {journal} {Applied Physics Letters}\ }\textbf {\bibinfo {volume} {112}},\ \bibinfo {pages} {262402} (\bibinfo {year} {2018})}\BibitemShut {NoStop}%
\bibitem [{Pt_()}]{Pt_chart}%
  \BibitemOpen
  \href {http:https://www.icdd.com/pdfsearch/} {\enquote {\bibinfo {title} {Pt crystallographic ref: pdf-2 {{ICDD}} file 00-004-0802},}\ }\BibitemShut {NoStop}%
\bibitem [{Co_()}]{Co_chart}%
  \BibitemOpen
  \href@noop {} {\enquote {\bibinfo {title} {Co crystallographic ref: pdf-2 {{ICDD}} file 00-015-0806},}\ }\BibitemShut {NoStop}%
\bibitem [{\citenamefont {Quinteros}\ \emph {et~al.}(2021)\citenamefont {Quinteros}, \citenamefont {{Cortés Burgos}}, \citenamefont {Albornoz}, \citenamefont {Gómez}, \citenamefont {Granell}, \citenamefont {Golmar}, \citenamefont {Ibarra}, \citenamefont {Bustingorry}, \citenamefont {Curiale},\ and\ \citenamefont {Granada}}]{QuinterosGrowth}%
  \BibitemOpen
  \bibfield  {author} {\bibinfo {author} {\bibfnamefont {C.~P.}\ \bibnamefont {Quinteros}}, \bibinfo {author} {\bibfnamefont {M.~J.}\ \bibnamefont {{Cortés Burgos}}}, \bibinfo {author} {\bibfnamefont {L.~J.}\ \bibnamefont {Albornoz}}, \bibinfo {author} {\bibfnamefont {J.~E.}\ \bibnamefont {Gómez}}, \bibinfo {author} {\bibfnamefont {P.}~\bibnamefont {Granell}}, \bibinfo {author} {\bibfnamefont {F.}~\bibnamefont {Golmar}}, \bibinfo {author} {\bibfnamefont {M.~L.}\ \bibnamefont {Ibarra}}, \bibinfo {author} {\bibfnamefont {S.}~\bibnamefont {Bustingorry}}, \bibinfo {author} {\bibfnamefont {J.}~\bibnamefont {Curiale}},\ and\ \bibinfo {author} {\bibfnamefont {M.}~\bibnamefont {Granada}},\ }\bibfield  {title} {\enquote {\bibinfo {title} {Impact of growth conditions on the domain nucleation and domain wall propagation in {{Pt}}/{{Co}}/{{Pt}} stacks},}\ }\href {https://doi.org/10.1088/1361-6463/abb849} {\bibfield  {journal} {\bibinfo  {journal} {Journal of Physics D: Applied Physics}\ }\textbf {\bibinfo {volume}
  {54}},\ \bibinfo {pages} {015002} (\bibinfo {year} {2021})}\BibitemShut {NoStop}%
\bibitem [{\citenamefont {Rossignol}\ and\ \citenamefont {Perrin}(2001)}]{rossignolPicosecondUltrasonicsStudy2001}%
  \BibitemOpen
  \bibfield  {author} {\bibinfo {author} {\bibfnamefont {C.}~\bibnamefont {Rossignol}}\ and\ \bibinfo {author} {\bibfnamefont {B.}~\bibnamefont {Perrin}},\ }\bibfield  {title} {\enquote {\bibinfo {title} {Picosecond ultrasonics study of periodic multilayers},}\ }\href {https://doi.org/10.14891/analscisp.17icpp.0.s245.0} {\bibfield  {journal} {\bibinfo  {journal} {Analytical Sciences/Supplements}\ }\textbf {\bibinfo {volume} {17}},\ \bibinfo {pages} {s245--s248} (\bibinfo {year} {2001})}\BibitemShut {NoStop}%
\bibitem [{\citenamefont {Thomsen}\ \emph {et~al.}(1986)\citenamefont {Thomsen}, \citenamefont {Grahn}, \citenamefont {Maris},\ and\ \citenamefont {Tauc}}]{thomsenSurfaceGenerationDetection1986}%
  \BibitemOpen
  \bibfield  {author} {\bibinfo {author} {\bibfnamefont {C.}~\bibnamefont {Thomsen}}, \bibinfo {author} {\bibfnamefont {H.~T.}\ \bibnamefont {Grahn}}, \bibinfo {author} {\bibfnamefont {H.~J.}\ \bibnamefont {Maris}},\ and\ \bibinfo {author} {\bibfnamefont {J.}~\bibnamefont {Tauc}},\ }\bibfield  {title} {\enquote {\bibinfo {title} {Surface generation and detection of phonons by picosecond light pulses},}\ }\href {https://doi.org/10.1103/PhysRevB.34.4129} {\bibfield  {journal} {\bibinfo  {journal} {Physical Review B}\ }\textbf {\bibinfo {volume} {34}},\ \bibinfo {pages} {4129--4138} (\bibinfo {year} {1986})}\BibitemShut {NoStop}%
\bibitem [{\citenamefont {Ortiz}\ \emph {et~al.}(2020)\citenamefont {Ortiz}, \citenamefont {Pastier}, \citenamefont {Rodriguez}, \citenamefont {{Priya}}, \citenamefont {Lemaitre}, \citenamefont {{Gomez-Carbonell}}, \citenamefont {Sagnes}, \citenamefont {Harouri}, \citenamefont {Senellart}, \citenamefont {Giesz}, \citenamefont {Esmann},\ and\ \citenamefont {{Lanzillotti-Kimura}}}]{ortizFiberintegratedMicrocavitiesEfficient2020}%
  \BibitemOpen
  \bibfield  {author} {\bibinfo {author} {\bibfnamefont {O.}~\bibnamefont {Ortiz}}, \bibinfo {author} {\bibfnamefont {F.}~\bibnamefont {Pastier}}, \bibinfo {author} {\bibfnamefont {A.}~\bibnamefont {Rodriguez}}, \bibinfo {author} {\bibnamefont {{Priya}}}, \bibinfo {author} {\bibfnamefont {A.}~\bibnamefont {Lemaitre}}, \bibinfo {author} {\bibfnamefont {C.}~\bibnamefont {{Gomez-Carbonell}}}, \bibinfo {author} {\bibfnamefont {I.}~\bibnamefont {Sagnes}}, \bibinfo {author} {\bibfnamefont {A.}~\bibnamefont {Harouri}}, \bibinfo {author} {\bibfnamefont {P.}~\bibnamefont {Senellart}}, \bibinfo {author} {\bibfnamefont {V.}~\bibnamefont {Giesz}}, \bibinfo {author} {\bibfnamefont {M.}~\bibnamefont {Esmann}},\ and\ \bibinfo {author} {\bibfnamefont {N.~D.}\ \bibnamefont {{Lanzillotti-Kimura}}},\ }\bibfield  {title} {\enquote {\bibinfo {title} {Fiber-integrated microcavities for efficient generation of coherent acoustic phonons},}\ }\href {https://doi.org/10.1063/5.0026959} {\bibfield  {journal} {\bibinfo  {journal}
  {Applied Physics Letters}\ }\textbf {\bibinfo {volume} {117}},\ \bibinfo {pages} {183102} (\bibinfo {year} {2020})}\BibitemShut {NoStop}%
\bibitem [{\citenamefont {{Lanzillotti-Kimura}}\ \emph {et~al.}(2011)\citenamefont {{Lanzillotti-Kimura}}, \citenamefont {Fainstein}, \citenamefont {Perrin},\ and\ \citenamefont {Jusserand}}]{lanzillotti-kimuraTheoryCoherentGeneration2011}%
  \BibitemOpen
  \bibfield  {author} {\bibinfo {author} {\bibfnamefont {N.~D.}\ \bibnamefont {{Lanzillotti-Kimura}}}, \bibinfo {author} {\bibfnamefont {A.}~\bibnamefont {Fainstein}}, \bibinfo {author} {\bibfnamefont {B.}~\bibnamefont {Perrin}},\ and\ \bibinfo {author} {\bibfnamefont {B.}~\bibnamefont {Jusserand}},\ }\bibfield  {title} {\enquote {\bibinfo {title} {Theory of coherent generation and detection of thz acoustic phonons using optical microcavities},}\ }\href {https://doi.org/10.1103/PhysRevB.84.064307} {\bibfield  {journal} {\bibinfo  {journal} {Physical Review B}\ }\textbf {\bibinfo {volume} {84}},\ \bibinfo {pages} {064307} (\bibinfo {year} {2011})}\BibitemShut {NoStop}%
\bibitem [{\citenamefont {{Pascual-Winter}}\ \emph {et~al.}(2012)\citenamefont {{Pascual-Winter}}, \citenamefont {Fainstein}, \citenamefont {Jusserand}, \citenamefont {Perrin},\ and\ \citenamefont {Lema{\^i}tre}}]{pascual-winterSpectralResponsesPhonon2012}%
  \BibitemOpen
  \bibfield  {author} {\bibinfo {author} {\bibfnamefont {M.~F.}\ \bibnamefont {{Pascual-Winter}}}, \bibinfo {author} {\bibfnamefont {A.}~\bibnamefont {Fainstein}}, \bibinfo {author} {\bibfnamefont {B.}~\bibnamefont {Jusserand}}, \bibinfo {author} {\bibfnamefont {B.}~\bibnamefont {Perrin}},\ and\ \bibinfo {author} {\bibfnamefont {A.}~\bibnamefont {Lema{\^i}tre}},\ }\bibfield  {title} {\enquote {\bibinfo {title} {Spectral responses of phonon optical generation and detection in superlattices},}\ }\href {https://doi.org/10.1103/PhysRevB.85.235443} {\bibfield  {journal} {\bibinfo  {journal} {Physical Review B}\ }\textbf {\bibinfo {volume} {85}},\ \bibinfo {pages} {235443} (\bibinfo {year} {2012})}\BibitemShut {NoStop}%
\bibitem [{Cob()}]{CobaltCo}%
  \BibitemOpen
  \href@noop {} {\enquote {\bibinfo {title} {Matweb, material property data, cobalt},}\ }\bibinfo {howpublished} {https://www.matweb.com/search/-datasheet.aspx?bassnum=AMECo00}\BibitemShut {NoStop}%
\bibitem [{\citenamefont {Raki{\'c}}\ \emph {et~al.}(1998)\citenamefont {Raki{\'c}}, \citenamefont {Djuri{\v s}i{\'c}}, \citenamefont {Elazar},\ and\ \citenamefont {Majewski}}]{rakicOpticalPropertiesMetallic1998}%
  \BibitemOpen
  \bibfield  {author} {\bibinfo {author} {\bibfnamefont {A.~D.}\ \bibnamefont {Raki{\'c}}}, \bibinfo {author} {\bibfnamefont {A.~B.}\ \bibnamefont {Djuri{\v s}i{\'c}}}, \bibinfo {author} {\bibfnamefont {J.~M.}\ \bibnamefont {Elazar}},\ and\ \bibinfo {author} {\bibfnamefont {M.~L.}\ \bibnamefont {Majewski}},\ }\bibfield  {title} {\enquote {\bibinfo {title} {Optical properties of metallic films for vertical-cavity optoelectronic devices},}\ }\href {https://doi.org/10.1364/AO.37.005271} {\bibfield  {journal} {\bibinfo  {journal} {Applied Optics}\ }\textbf {\bibinfo {volume} {37}},\ \bibinfo {pages} {5271} (\bibinfo {year} {1998})}\BibitemShut {NoStop}%
\bibitem [{\citenamefont {Polyanskiy}(2024)}]{polyanskiyRefractiveindexinfoDatabaseOptical2024}%
  \BibitemOpen
  \bibfield  {author} {\bibinfo {author} {\bibfnamefont {M.~N.}\ \bibnamefont {Polyanskiy}},\ }\bibfield  {title} {\enquote {\bibinfo {title} {Refractiveindex.info database of optical constants},}\ }\href {https://doi.org/10.1038/s41597-023-02898-2} {\bibfield  {journal} {\bibinfo  {journal} {Scientific Data}\ }\textbf {\bibinfo {volume} {11}},\ \bibinfo {pages} {94} (\bibinfo {year} {2024})}\BibitemShut {NoStop}%
\bibitem [{\citenamefont {Johnson}\ and\ \citenamefont {Christy}(1974)}]{johnsonOpticalConstantsTransition1974}%
  \BibitemOpen
  \bibfield  {author} {\bibinfo {author} {\bibfnamefont {P.}~\bibnamefont {Johnson}}\ and\ \bibinfo {author} {\bibfnamefont {R.}~\bibnamefont {Christy}},\ }\bibfield  {title} {\enquote {\bibinfo {title} {Optical constants of transition metals: {{Ti}}, {{V}}, {{Cr}}, {{Mn}}, {{Fe}}, {{Co}}, {{Ni}}, and {{Pd}}},}\ }\href {https://doi.org/10.1103/PhysRevB.9.5056} {\bibfield  {journal} {\bibinfo  {journal} {Physical Review B}\ }\textbf {\bibinfo {volume} {9}},\ \bibinfo {pages} {5056--5070} (\bibinfo {year} {1974})}\BibitemShut {NoStop}%
\bibitem [{\citenamefont {{Cardozo de Oliveira}}\ \emph {et~al.}(2023)\citenamefont {{Cardozo de Oliveira}}, \citenamefont {Xiang}, \citenamefont {Esmann}, \citenamefont {Lopez~Abdala}, \citenamefont {Fuertes}, \citenamefont {Bruchhausen}, \citenamefont {Pastoriza}, \citenamefont {Perrin}, \citenamefont {{Soler-Illia}},\ and\ \citenamefont {{Lanzillotti-Kimura}}}]{cardozodeoliveiraProbingGigahertzCoherent2023}%
  \BibitemOpen
  \bibfield  {author} {\bibinfo {author} {\bibfnamefont {E.~R.}\ \bibnamefont {{Cardozo de Oliveira}}}, \bibinfo {author} {\bibfnamefont {C.}~\bibnamefont {Xiang}}, \bibinfo {author} {\bibfnamefont {M.}~\bibnamefont {Esmann}}, \bibinfo {author} {\bibfnamefont {N.}~\bibnamefont {Lopez~Abdala}}, \bibinfo {author} {\bibfnamefont {M.~C.}\ \bibnamefont {Fuertes}}, \bibinfo {author} {\bibfnamefont {A.}~\bibnamefont {Bruchhausen}}, \bibinfo {author} {\bibfnamefont {H.}~\bibnamefont {Pastoriza}}, \bibinfo {author} {\bibfnamefont {B.}~\bibnamefont {Perrin}}, \bibinfo {author} {\bibfnamefont {G.~J. A.~A.}\ \bibnamefont {{Soler-Illia}}},\ and\ \bibinfo {author} {\bibfnamefont {N.~D.}\ \bibnamefont {{Lanzillotti-Kimura}}},\ }\bibfield  {title} {\enquote {\bibinfo {title} {Probing gigahertz coherent acoustic phonons in {{TiO}}$_2$ mesoporous thin films},}\ }\href {https://doi.org/10.1016/j.pacs.2023.100472} {\bibfield  {journal} {\bibinfo  {journal} {Photoacoustics}\ }\textbf {\bibinfo {volume} {30}},\ \bibinfo {pages}
  {100472} (\bibinfo {year} {2023})}\BibitemShut {NoStop}%
\bibitem [{\citenamefont {{Lanzillotti-Kimura}}\ \emph {et~al.}(2018)\citenamefont {{Lanzillotti-Kimura}}, \citenamefont {O'Brien}, \citenamefont {Rho}, \citenamefont {Suchowski}, \citenamefont {Yin},\ and\ \citenamefont {Zhang}}]{lanzillotti-kimuraPolarizationcontrolledCoherentPhonon2018}%
  \BibitemOpen
  \bibfield  {author} {\bibinfo {author} {\bibfnamefont {N.~D.}\ \bibnamefont {{Lanzillotti-Kimura}}}, \bibinfo {author} {\bibfnamefont {K.~P.}\ \bibnamefont {O'Brien}}, \bibinfo {author} {\bibfnamefont {J.}~\bibnamefont {Rho}}, \bibinfo {author} {\bibfnamefont {H.}~\bibnamefont {Suchowski}}, \bibinfo {author} {\bibfnamefont {X.}~\bibnamefont {Yin}},\ and\ \bibinfo {author} {\bibfnamefont {X.}~\bibnamefont {Zhang}},\ }\bibfield  {title} {\enquote {\bibinfo {title} {Polarization-controlled coherent phonon generation in acoustoplasmonic metasurfaces},}\ }\href {https://doi.org/10.1103/PhysRevB.97.235403} {\bibfield  {journal} {\bibinfo  {journal} {Physical Review B}\ }\textbf {\bibinfo {volume} {97}},\ \bibinfo {pages} {235403} (\bibinfo {year} {2018})}\BibitemShut {NoStop}%
\end{thebibliography}%

\end{document}